\newif\iffullversion
\fullversionfalse

\documentclass[review]{elsarticle}
\usepackage[cmex10]{amsmath}
\usepackage{array}

\usepackage{fixltx2e}
\usepackage{stfloats}
\usepackage{url}
\usepackage{graphicx,color}
\usepackage{caption}
\usepackage{subcaption}

\usepackage{epsfig}
\usepackage{epstopdf}
\usepackage{stys/mathfunc}
\usepackage{stys/compress}
\usepackage{stys/ertemcommon} 
\usepackage{stys/ertemtext} 
\usepackage{stys/ertemalgo} 

\definecolor{orange}{RGB}{255,127,0}
\newcommand{\drafta}[1]{ {\color{black} #1}}

\journal{Future Generation Computer Systems}

\title{Auditable Versioned Data Storage Outsourcing}

\begin{document}
\author[add1]{Ertem Esiner}
\ead{ertem001@ntu.edu.sg}
\author[add1]{Anwitaman Datta}
\ead{anwitaman@ntu.edu.sg}

\address[add1]{School of Computer Engineering\\
Nanyang Technological University\\
Singapore\\}

\begin{abstract}
Auditability is crucial for data outsourcing, facilitating accountability and identifying data loss or corruption incidents in a timely manner, reducing in turn the risks from such losses. In recent years, in synch with the growing trend of outsourcing, a lot of progress has been made in designing probabilistic (for efficiency) provable data possession (PDP) schemes. However, even the recent and advanced PDP solutions that do deal with dynamic data, do so in a limited manner, and for only the latest version of the data. A naive solution treating different versions in isolation would work, but leads to tremendous overheads, and is undesirable. In this paper, we present algorithms to achieve full persistence (all intermediate configurations are preserved and are modifiable) for an optimized skip list (known as \flexlist) so that versioned data can be audited. The proposed scheme provides deduplication at the level of logical, variable sized blocks, such that only the altered parts of the different versions are kept, while the persistent data-structure facilitates access (read) of any arbitrary version with the same storage and process efficiency that state-of-the-art dynamic PDP solutions provide for only the current version, while commit (write) operations incur around 5\% additional time. Furthermore, the time overhead for auditing arbitrary versions in addition to the latest version is imperceptible even on a low-end server. Additionally, the application of our approach opens up the possibility to naturally support block level deduplication. While a naive solution to audit versions would copy the whole data and the data structure for each version, our solution utilises storage space amounting very close to the most efficient delta-based solutions. Accordingly, we explore how the proposed data structure benefits the system with block level deduplication besides adding auditability property, and how it can be integrated with a state-of-the-art versioning system (Git), and in the process scale the storage efficiency of Git, and thus help scale the size of data to be stored in Git, without compromising the retrieval efficiency of arbitrary versions. 

\end{abstract}

\maketitle
\iffullversion
\begin{center}
\end{center}
\fi

\section{Introduction}\label{sec:intro}

One of the natural challenges in outsourcing data to third party services is to ensure its integrity over time. This has led to the study of  probabilistic techniques for Provable Data Possession (PDP) \cite{PDP}, in order to efficiently determine the loss or corruption of (substantial subset) of the outsourced data. Early works on PDP \cite{POR,compactPOR,generalPOR} dealt with static data, limiting their practicality. This in turn led more recently to the exploration of mechanisms better suited for dynamic data, e.g., DPDP \cite{DPDP}, FlexDPDP \cite{blind} and Dynamic POR using Oblivious RAM \cite{PORAM}. These later solutions do not consider provability of the intermediate versions, but instead focus on the final/latest version. This leads to the natural question on whether the meta-information maintained to carry out dynamic proof of data possession can be amortized to also carry out version control, and how in turn to adapt the data-structures and algorithms to carry out data possession and integrity verification while supporting the various versions of the data. The presented work targets precisely this sweet spot, which we believe paves way for deployment of wider class of applications using an outsourced backend storage service.

An obvious way to go about is to apply PDP separately over a version maintenance mechanism, e.g., as in \cite{skipdeltaChen}. However, by instead adapting a dynamic PDP solution, we can amortize the data structures, and realize versioning with marginal further overheads. Specifically, given its ability to efficiently support mutable content, albeit only for the latest version, we leverage on the FlexDPDP \cite{blind} approach which uses a probabilistically balanced data structure allowing for variable sized blocks, giving flexibility to accommodate arbitrary changes. 

In order to support the versioning of the data, we need a data structure which is fully persistent. So to say, the data structure should be such that it should be possible to access and modify any version of the data \cite{Driscoll198986}, possibly creating version forks in the process. For answering audit queries the data structure needs to be authenticated (each node keeps a hash value), yet, for practicality, the data-structure also needs to be efficient. The original skip list data structure was optimised in \cite{DPDP} and \cite{blind}, resulting in the use of only half the number of nodes and links above level 0 and a new traversal mechanism which allows the data structure to be used for PDP with dynamic data. The optimization diminishes the number of most costly operation (namely hashing) and stored hash values.

The prior works however delve in ephemeral data structures - any update in the data structure results in the loss of the previous version. In this work we present necessary algorithms to realize full persistence to support multiple versions of data. Our experiments that the performance overheads - for either data manipulation (read/write of arbitrary version) or for auditing - are barely visible with respect to the underlying ephemeral data structure that supported only the latest version of dynamic data. Additionally, we discuss how our approach can be integrated with Git \cite{gitt} version control system, and in doing so, in addition to the provable data possession guarantee, our approach derives significant savings in storage by facilitating block-level deduplication, without compromising data access performance.

The contributions, \drafta{and consequent advantages and key performance indicators of the work presented in this} paper are as follows:\\
 \noindent $\blacksquare$ We propose algorithms to realize persistence of a data structure (FlexList) (Section \ref{sec:ver_alg}), and thus extend its existing functionality from supporting the audit of the latest version of a mutable content to that of arbitrary or all versions of the mutable data. Our approach is general enough and can be adapted for optimized skip list and rank-based skip list with slight changes to the traversal functions (namely canGoBelow and canGoAfter).
 
 \noindent $\blacksquare$ Our versioning algorithms add only 5\% computational overheads to the original \cite{DPDP,blind} schemes when committing a new version of data.

 \noindent $\blacksquare$ Computational overhead for challenging versions is in the worst case $log(v)$, where $v$ is the number of versions. For a 1GB of file with 1500 versions, our approach adds only a millisecond more latency for auditing the versions with respect to approaches which could store and audit only the last version, even with a low-end server.

\noindent  $\blacksquare$ Each version of the data is retrievable with $O(n)$ I/O operations, where $n$ is the number of file system blocks (fixed sized, determined by the underlying file system) of the data (assuming that the data structure is in the main memory).

 \noindent $\blacksquare$ Unlike existing delta based versioning solutions (such as SVN\cite{svnn}), each byte of each version can be reached by expected worst-case of $O(log(v) + log(b))$ traversal on the data structure, where $v$ and $b$ are number versions and blocks respectively. And unlike all tree-based solutions it achieves this regardless of the alteration pattern of the user and without needing a re-balancing operation. 
 
 \noindent $\blacksquare$ Thus, our approach opens up the possibility to realize block level deduplication across versions of data, without compromising the efficiency of access or manipulation of any of the versions. We expose this by exploring integration of our solution with a state of the art versioning system, Git.

\section{Related Work}\label{sec:rel}

The related works can be broadly classified across two different aspects: (i) mechanisms for keeping track of mutable content and (ii) auditing outsourced data. 
The former encompasses a wide spectrum of services, including system backups, typically based on snapshots or copy on write mechanisms \cite{Hitz1994,Chutani92theepisode,Johnson96overviewof,Kistler1992,Lee96petal,Peterson03ext3cow,Quinlan91acached,Quinlan2002,Roome91,tops20U,McCoy1055811,MuniswamyReddy2004,Soules2003mev}, and version management systems such as SVN \cite{svnn} and Git \cite{gitt}. The techniques we propose are applicable for storing the mutable content arising from either setting, however, in the current work we do not address synchronization and conflict resolution, and those additional mechanisms would have to be integrated on top in order to realize a full-fledged version management system. We instead discuss how our approach complements a full-fledged version control framework like Git \cite{gitt}. The primary focus of this work is thus only on the aspect of storing and manipulating the changed content, and the persistent data structures to efficiently access them.  

The other aspect falls under the broader umbrella of security. While confidentiality of the data - be it in transit or when it is stored - can readily be achieved with encryption, the current version control systems implicitly assume a trusted storage service as far as data corruption or loss is concerned. One of the challenges of outsourcing is however precisely to ascertain integrity of the stored content. That's the notion of auditing.  

The pioneering work on provable data possession for auditing \cite{PDP}, and several subsequent works \cite{POR,compactPOR,generalPOR,PORAM} support append-only changes to the data. These were followed by other works \cite{wangPOR} using a Merkle Hash Tree and other balanced data structures such as range-based 2-3 tree \cite{authenticated23trees,FDPOR}, which however suffer from implementation and performance complexities, particularly necessitated by re-balancing operations. 
		
In \cite{DPDP}, an optimized skip list was proposed with modified traversal algorithm, namely the `rank-based authenticated skip list', and provided the first dynamic provable data possession scheme. It was extended in FlexDPDP \cite{blind} to accommodate variable block sizes. A radically different approach provides dynamic proof of retrievability using Oblivious RAM \cite{PORAM}, and achieves stronger guarantees but is not as efficient as \cite{DPDP,blind} solutions. All these approaches keep track of only the latest version, and hence, as it is, inadequate for versioning, or proof of data possession of multiple versions of the data. A naive solution will be to keep multiple instances of such a data structure and the corresponding data separately, one instance per version, but such an approach is clearly inefficient. We will explore how the data and data structure across versions can be superimposed together, thereby keeping track of only the difference across versions, and yet, keeping data access and manipulation efficiency comparable to storing each version separately, and not suffer the qualms of delta-based versioning solutions.

To understand this further, we next switch to the discussion of some popular versioning mechanisms, and the implications of our work with respect to these. SVN (Subversion) \cite{svnn} is a delta-based versioning mechanism. A delta represents the difference from the previous version. Thus, an SVN server stores the initial file and deltas for subsequent commits. An easy approach will be to use a static solution for the initial data and each delta as in \cite{skipdeltaChen}. The audit in this setting is fast. On the other hand, not only retrieving the later versions is slow (needs to fetch and apply all the deltas) but also, the earlier versions cannot be removed since \cite{PDP} doesn't allow any insertion or removal, but only append operations. Therefore a dynamic solution is required to support all sorts of mutations. The way the deltas are stored comes into prominence. With our approach, one can achieve fast retrieval as in Git (discussed next) and less storage overhead as in SVN, while providing data possession guarantees.
	
Git \cite{gitt} overcomes the reconstruction inefficiency problem of the delta-based solutions. It stores a directory of data in blobs where a blob consists of the content of a file. Each blob is indexed by a tree structure. Whenever a change occurs inside specific files (a commit), new corresponding blobs and a new tree is created for the commit operation. Therefore all of the versions are at $log(n)$ distance when requested, where $n$ is the number of different files (blobs). Blobs corresponding to the unchanged files are reused across versions, thereby achieving deduplication at file level, nevertheless, for large files, a lot of storage space is wasted whenever there is even a small change. With our approach, the deduplication is moved to a finer granularity of data blocks, reducing storage overheads when storing multiple versions of data. Leaving the indexing data structure as a tree (where we suggest a Merkle hash tree), we apply our solution to file level. 

We chose \flexlist\ as the underlying data structure to design our solution since other dynamic provable data possession techniques such as Zang et al's update tree \cite{Zhangutreedpdp} or red-black tree \cite{bung1998authenticated} need re-balancing, while \flexlist\ keeps itself probabilistically balanced, inheriting an intrinsic property of skip list which it enhances. This eliminates the need for re-balancing, which is important for an authenticated data structure, since re-balancing is very expensive since it requires many hash calculations.

\section{Preliminaries: FlexList}

\begin{figure}[htb]
\begin{center}
	\includegraphics[scale = 0.8]{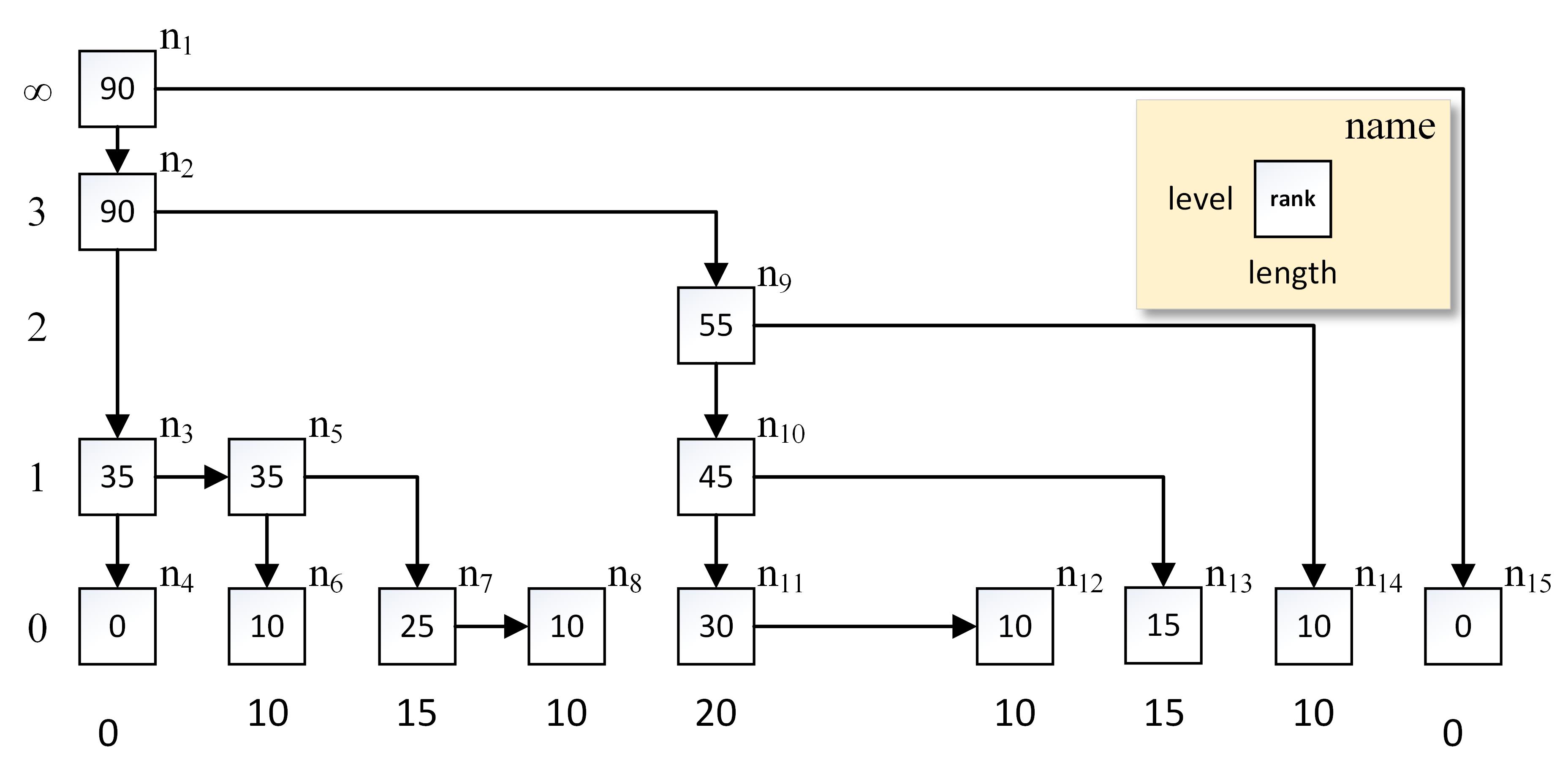}
\caption{A \flexlist\ example.}\label{fig:ToyFlexList}
\end{center}
\end{figure}

In this section, we provide a summary of prior arts, specifically on \flexlist\ (see Figure \ref{fig:ToyFlexList}) which realizes a probabilistically balanced authenticated dictionary enabling provable data possession of variable block sized dynamic data, by extending in turn a rank-based authenticated skip list \cite{DPDP}. 

We explain how the \flexlist\ works using examples. In Figure \ref{fig:ToyFlexList}, consider node n$_{2}$ which keeps two outgoing links, a rank value (90) and a level value (3). By following the links we reach n$_{3}$ and n$_{9}$, called the below and after nodes respectively. The rank is the value that represents how many bytes of the real data can be reached passing through that particular node. The rank value is computed by adding rank values of the node below and the node after (we take 0 as the rank of after if the link is null). If the below link is null (at leaf level) the length of the corresponding data piece is used. The level represents the height of the node in the data structure. Last but not least, each node keeps a hash value generated using a collision-resistant hash function (such as SHA1) which uses all the above as inputs, namely; hash values of the below and after nodes, the rank value and the level value. ß

Now, consider n$_{7}$ which has an after link to n$_{8}$, but doesn't have a below link. A leaf level node has a pointer to the data block that it corresponds. In the figure the values shown below the nodes represent the lengths of the data blocks. We use the hash value of the after node (if there is none, then null), the information of the data block, the level and the rank values of the block. For simplicity, we can think the information of the data block as the block content. In the real protocols, to achieve the same level of security with a lesser communication cost, a value named Tag is used \cite{DPDP}. Note that the \flexlist\ has 2 sentinel nodes (n$_{4}$ and n$_{15}$). These nodes don't provide any new dependencies and they are not related with any data. The sole reason of their existence is to make the algorithms and the figures more lucid. Since the hash value of the root node of the \flexlist\ depends on each block and each node on the data structure, making \flexlist\ an authenticated dictionary.

\subsection{\textbf{A Search Example}}
Say we are searching the block that includes the byte with index 40. We start with the root node n$_{1}$ and we check the below node's rank (note that the rank value shows how many bytes can be reached from that node). Since, 40 $<$ 90 we conclude that we should follow the below link and reach n$_{2}$. Repeating the same check we see that 40 $>$ 35 therefore we need to follow the after link this time, and arrive at n$_{9}$. When we follow an after link, we leave behind some bytes on the left that can not be reached anymore. We deduct those bytes from the index we are looking for. We continue our search with index 40 - 35 (rank of n$_{3}$) = 5. Following the same rules as the first one we reach n$_{10}$ and n$_{11}$ respectively. Knowing the length of the corresponding data for n$_{11}$ is 20, we stop since index 5 can be reached at that point.

\begin{figure}[htb]
\begin{center}
	\includegraphics[scale = 0.8]{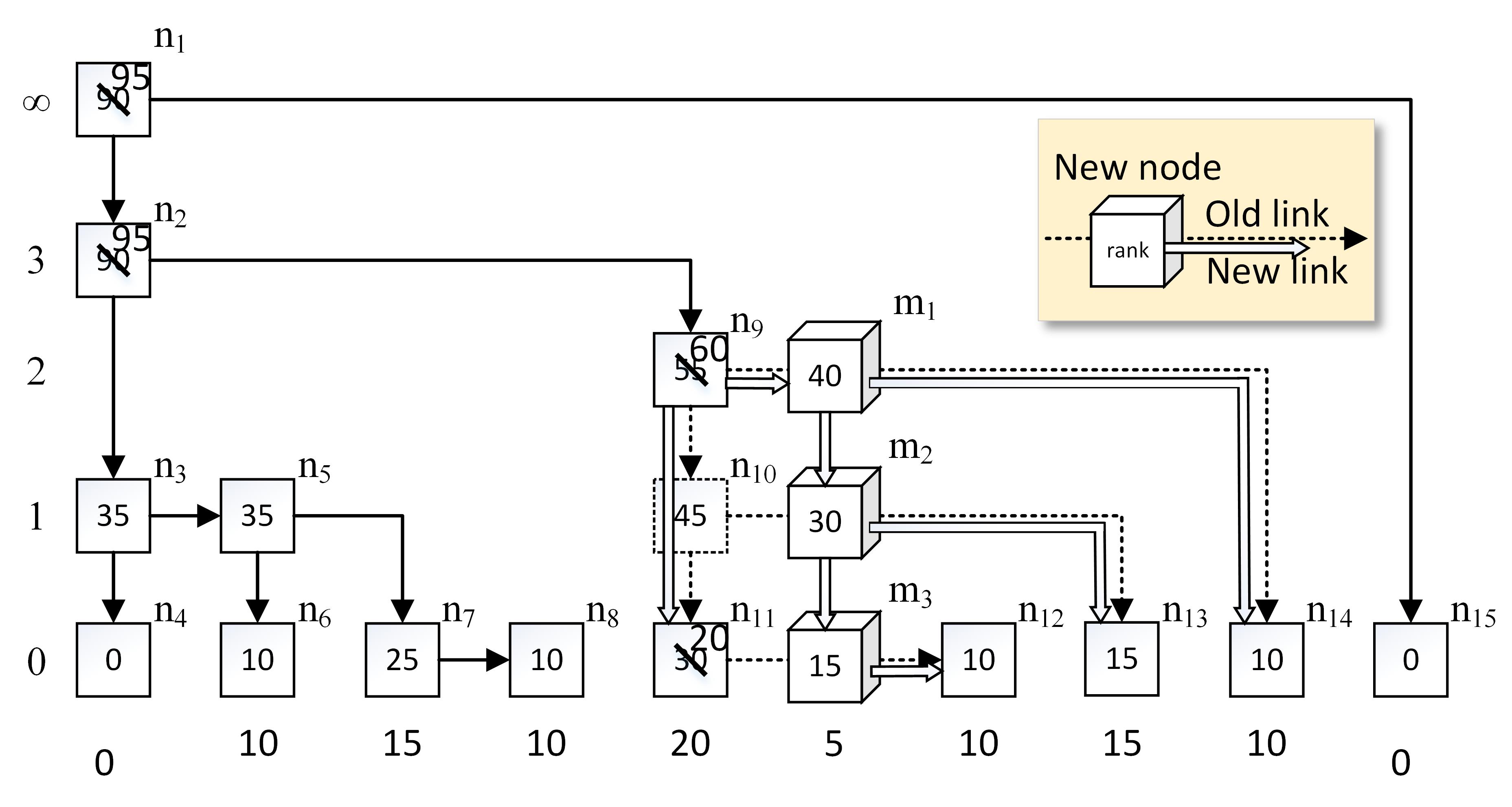}
\caption{An \textit{insert} example on the \flexlist .}
\label{fig:ToyInsertFlexList}
\end{center}
\end{figure}
\subsection{\textbf{An Insert and Remove Example}}\label{text:insert}

Insertion and removal to a \flexlist\ are a bit more complicated. We must preserve optimality of the data structure by removing nodes which become unnecessary after the operation and by creating new nodes if a connection point is empty. Optimality is where we have links and nodes if they are creating new dependencies otherwise we call them unnecessary. Lets say in Figure \ref{fig:ToyInsertFlexList}, we insert a data block of length 5 to index 55.

\textbf{Insertion:}  A level is chosen for the new node randomly, following a geometric distribution with parameter 0.5. The expected depth of the resulting data structure is $log(b)$ where $b$ is the number of leaf level nodes (number of blocks), ensuring that the skip list (thus the \flexlist ) is probabilistically balanced and doesn't need re-balancing \cite{PughCookbook}.

We start from the root n$_{1}$ and with the same rule as in search we follow the links until we reach the node before the first insertion point (n$_{9}$ in our Example). Then, we create m$_{1}$ and connect it between n$_{9}$ and n$_{14}$. We continue following the links and reach n$_{10}$ and stop since n$_{10}$ has an after link which crosses our new nodes index. We create m$_{2}$, connect it to m$_{1}$ and copy the after link from n$_{10}$ to m$_{2}$. As the node n$_{10}$ loses its after node it becomes unnecessary because it is dependent to none but only n$_{11}$. Then we follow below link of n$_{10}$ and reach n$_{11}$ from which we will copy the after link. We create m$_{3}$, connect it to m$_{2}$ and copy after link from n$_{11}$ to m$_{3}$. Since n$_{11}$ is a leaf level node, we do not remove it since it is our path to the corresponding data block. We calculate all the rank and hash values of the nodes on the path, backwards until the root, to finalize the operation.

\drafta{\textbf{Removal:} Simply put, the remove operation is reverse of the insertion operation. In the running example, we are to remove node with index 55 from Figure \ref{fig:ToyInsertFlexList}, and obtain back the FlexList in Figure \ref{fig:ToyFlexList}.}

\subsection{Pre-processing for Data Possession Guarantees}	
Pre-processing is done at both the client and the server sides. We show the concept as if there is only one file at the server for simplicity. We discuss file hierarchy later. For the randomness (refer to \ref{text:insert}) of the data structure, they both use the same seed to construct the \flexlist\ over the data divided into pieces. Another choice would be the client to construct the data structure and send it to the server. This saves processing time for the server but adds communication cost. When both parties have the data structure, the hash value of the root is the meta-data of the client for later integrity checks. In our scheme, 
the root of the layer 2 data structure is the meta data to the client. We use layer 2 to keep each version's root, meaning after each commit applied to layer 1 \flexlist\ the newly created root for the new version is to be added to layer 2. Note that the client's space complexity for meta-data is still $O(1)$. Initially after a pre-process the root of the layer 1 \flexlist\ is the first and only element of the layer 2 data structure. 

\section{Auditing with Versioning}\label{sec:ver}

\begin{figure}
\begin{center}
	\includegraphics[scale = 0.65]{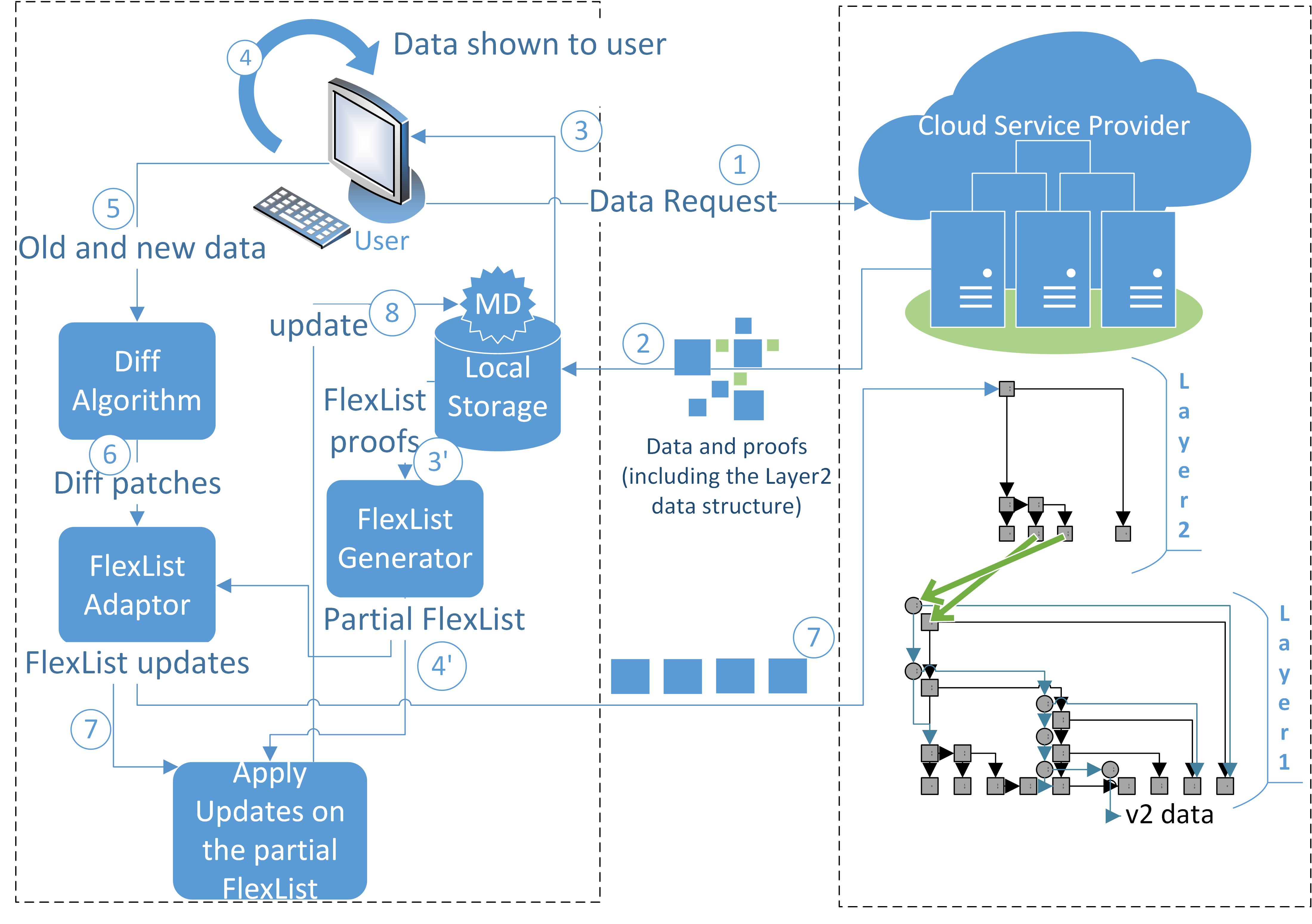}
\caption{Illustration of a commit operation of a new version for one file.}\label{fig:mainDiagram}
\end{center}
\end{figure}

In this section, we discuss versioning. We first present a high level illustration (Figure \ref{fig:mainDiagram}) of the modified update phase of the FlexDPDP scheme that incorporates the versioning feature. Then we provide the operations to persistently modify, insert and remove content in order to keep track of the modified data (used at both the server and the client whenever an update occurs). We then discuss how challenges over the versions and roll-backs work. Last but not least, we explore how our approach's integration benefits a state-of-the-art versioning system (Git).

\subsection{Update Phase}
After the pre-processing, both sides (client and server) have the exact same data structures. Later when the client wants to make any change she will take the following steps, illustrated in Figure \ref{fig:mainDiagram}. 
In the Figure \textbf{`\flexlist\ Generator'} stands for the algorithm which creates a partial \flexlist\ using a proof. Since a proof consists of all nodes to be altered, the partial \flexlist\ includes all of the nodes that we need for an update.  \textbf{`\flexlist\ Adaptor'} stands for the algorithm that translates standard diff output (such as [insert "in red w" to index 59]) to data structure operations such as [modify the block starting with index 50 with "the lady in red was"], or something like [insert the block starting with index 59 with content "n red w"].

When a user wants to interact with the remotely stored data, the client requests data from the server (Step 1) that the client wants to work on. The server's response includes the requested part of the data along with a DPDP proof (Step 2). The proof consists of the necessary information from the two data structures 
to compute the meta-data kept at the client. Therefore, the client may check the integrity of the received parts that the client will work on. We do not include verification step in the illustration for simplicity. While the data is sent to the user and the user interacts with the data (Steps 3 and 4), the proof is processed by the \flexlist\ Generator and the output (a partial \flexlist) awaits the updates by the user (Steps 3' and 4'). When the client is done updating, the resulting changes are parsed by a diff algorithm (Step 5) and sent to the \flexlist\ Adaptor, where they are turned into block updates (Step 6) which consists of modify, insert and remove operations. Then the updates are applied to the partial \flexlist\ at the client side while they are being sent to the server for application to the remotely stored data (Step 7). The client needs to apply the block updates to the partial FlexList of layer 1 and add the resulting root value to the data structure of layer 2 to compute her new meta-data. Note that partial \flexlist\ along with the updates from the client includes all necessary nodes to compute the new root of the layer 2 data structure.  After the client computes and stores the new meta-data (Step 8), she can remove the partial \flexlist\ and all the update information. 

\begin{table}
\begin{tabular}{m{2cm}|l|m{1.5cm}|m{1.5cm}|m{1.5cm}|}
\cline{2-5}
                                                     & PDP  & Key-based Skip List & Rank-based Skip List & FlexList \\ \hline
\multicolumn{1}{|l|}{Code Re-use}                    & no   & yes*                & yes                  & yes      \\ \hline
\multicolumn{1}{|l|}{Branching needs new algorithms} & yes  & no                  & yes                  & yes      \\ \hline
\multicolumn{1}{|l|}{Process time for challenges}    & $O(1)$ & $O(log\;c)$            & $O(log\;c)$             & $O(log\;c)$ \\ \hline
\multicolumn{1}{|l|}{Removing middle versions}    & no   & yes                 & yes                  & yes      \\ \hline
\multicolumn{1}{|l|}{Removing a whole branch}        & $O(v)$ & $O(v\;log\;c)$             & $O(v\;log\;c)$              & $O(v\;log\;c)$ \\ \hline
\multicolumn{1}{|m{3cm}|}{Needs extra field to keep \#blocks effected in a version 
}   &   yes   &  yes   &       yes    &  no        \\ \hline


\end{tabular}
\caption{Level 2 structure choice and properties.}
\label{tab:ds}
\end{table}

\drafta{There are several alternatives to choose from to serve as the layer 2 data structure. Main options are presented in Table \ref{tab:ds}. In the table, $v$ and $c$ are number of versions and commits respectively.}

\drafta{Assuming each update is of size 20KB (which is a fairly big size for an average update) and we challenge 460 different nodes (for detecting data loss of 1\% with 99\% confidence \cite{PDP}) which alter from a version to another, the overheads of challenging versions while using the different layer 2 structure choices are as follows. Time consumed by skip list, rank-based skip list and FlexList are approximately the same, and are around 0.65ms for 1000 commits, 0.79ms for 5000 commits, 0.9ms for 10000 commits and 1.19ms for 50000 commits, while the time consumed by PDP is around 0.53ms irrespective of the number of commits. Note that we used an altered version of PDP here for a fair comparison, where we mitigate server process by transferring it to the client, similar to as DPDP solutions do.}

\drafta{Though PDP is faster than the other three alternatives, it has other limitations (as tabulated). Among the remaining three options, there is no discernible advantage of any one approach over the others, and any of the three would be suitable as a choice of layer 2 data structure. For the rest of the exposition of our approach and the experiments, we chose FlexList.}

\subsection{Algorithms for Versioning}\label{sec:ver_alg}

There are several ways of versioning. First and the very basic one is to keep each version of the data. This solution requires no query overhead when a previous version is needed and any static provable data possession solution may work on this case by processing each version as a new data stored. On the other hand this highly increases the space requirements and the time spent for the updates. Another solution would be delta versioning which is recording updates and keep only the last or first version of the data. In this case, updates and the queries for the chosen version (first or last) are fast, it uses minimum space but the queries for the other versions are too slow. For this case a static PDP solution can be employed for the deltas (the update records) for the the scenario that only the first version is kept and all of the updates are kept as records. This seems inefficient since users tend to use the latest version. For a dynamic data, keeping the last version and having the PDP solution for the deltas doesn't work since we can not apply PDP to the latest version due to the fact that the PDP solution is efficient only under static setting. Therefore a dynamic PDP solution is needed to be employed for this case. In this Section we show how not to suffer slow queries for the past versions by presenting a path copying solution for the \flexlist . We summarize the crucial symbols used in the presented algorithms in table \ref{tab:symbols}. 

\begin{table}
\footnotesize
\centering
\begin{tabular}{| p{1.9cm} | p{9cm} |}
\hline
Symbol & Description \\ \hline
\textit{(new)}\CurrentNode\ & (new) current node \\ \hline
\PreviousNode\ & previous node, indicates the last node that current node moved from \\ \hline
\MissingNode\ & missing node, created when there is no node at the point where a node has to be linked\\ \hline
\NewNode\ & new node \\ \hline
\DeleteNode\ & node to be deleted \\ \hline
\After\ & the after neighbor of a node \\ \hline
\Below\ & the below neighbor of a node \\ \hline
\Rank\ & rank value of a node \\ \hline
\Index\ & index of a byte\\ \hline
\Npi\ & a boolean which is always true except in the inner loop of \textit{insert} algorithm \\ \hline 
\Stack\ & stack (initially empty), filled with all visited nodes during $search$, $modify$, $insert$ or $remove$ algorithms \\ \hline
\textit{canGoBelow}, \textit{canGoAfter} & functions that make decisions according to \flexlist\ mechanism \\ \hline
\textit{createMissing Node}& function that creates a node when a link needs to be connected to a point of another link, where there is no node present \\ \hline
\end{tabular}
\caption{Symbol descriptions of \flexlist\ algorithms (we reuse the notations from \cite{blind}).}
\label{tab:symbols}
\end{table}

\begin{algorithm}
{\fontsize{\AlgFontSize}{\AlgSpace}\selectfont
\SetKwComment{Comment}{}{}

\KwIn{\PreviousNode, \CurrentNode, \Index, \Level, $npi$, \NewCurrentNode, \Version}
\KwOut{\PreviousNode, \CurrentNode, \Index, \Stack, \NewCurrentNode}
\BlankLine

\Indp
    \Stack\ = new empty Stack \;
    \While{\CurrentNode\ \textnormal{can go} \Below\ \textsc{or} \After}{
        \PreviousNode\ = \NewCurrentNode\;
        \eIf{ \CanGoBelow(\CurrentNode, \Index) \textsc{and}  \CurrentNode.\Below.\NodeLevel\ $ \ge $ \Level\ \textsc{and} $ npi $ }{
            \CurrentNode\ =  \CurrentNode.\Below\;
            \underline{\textbf{\NewCurrentNode.\Below = new Node(\CurrentNode, \Version)\;}}
            \underline{\textbf{\NewCurrentNode = \NewCurrentNode.\Below\;}}
        }
        {	\If { \CanGoAfter(\CurrentNode, \Index) \textsc{and} \CurrentNode.\After.\NodeLevel\ $ \ge $ \Level}{
            \Index\ =  \Index\ - \CurrentNode.\Below.\Rank;
            \CurrentNode\ = \CurrentNode.\After\;
           \underline{\textbf{\NewCurrentNode.\After = new Node(\CurrentNode, \Version)\;}}
           \underline{\textbf{\NewCurrentNode = \NewCurrentNode.\After\;}}
        	}
        }
        \underline{\textbf{add \NewCurrentNode\ to \Stack;}}
    }

\Indm
\caption{nextPos Algorithm}
\label{alg:nextPos}
}
\end{algorithm}

We first describe the \textit{nextPos} function (Algorithm \ref{alg:nextPos}) to move from a current node to a desired node of interest. Let us revisit the insert example of Figure \ref{fig:ToyInsertFlexList}. \textit{nextPos} starts from the root n$_{1}$ and moves the pointer to n$_{2}$, n$_{9}$, then stops. After the \textit{insert} creates m$_{1}$, \textit{nextPos} is called again to arrive at n$_{10}$. After the creation of m$_{2}$, \textit{nextPos} moves the pointer to n$_{11}$. As observed, the function covers the affected nodes on the path. By using this property we modify the algorithm as follows.

 We add two more inputs which are a pointer to a newly created node for versioning and the version number. By this function, we handle the main versioning of the manipulated nodes by creating version nodes each time we move the pointer of the current node. We create the version node by copying the current node and setting the version number and move the pointer of the new current node (namely newcn in the algorithms) to the newly created one. When traversal ends and some specific operations are required, we handle them in the main algorithms that we present below. Last but not least, our \textit{nextPos} algorithm pushes all the newly created nodes to the stack (Line 11), since all these nodes are on the search path and will need a recalculation according to main FlexList algorithms \cite{blind}.

	
	\subsubsection{Modify Operation}
	
	\begin{figure}
	\begin{center}
		\includegraphics[scale = 0.8]{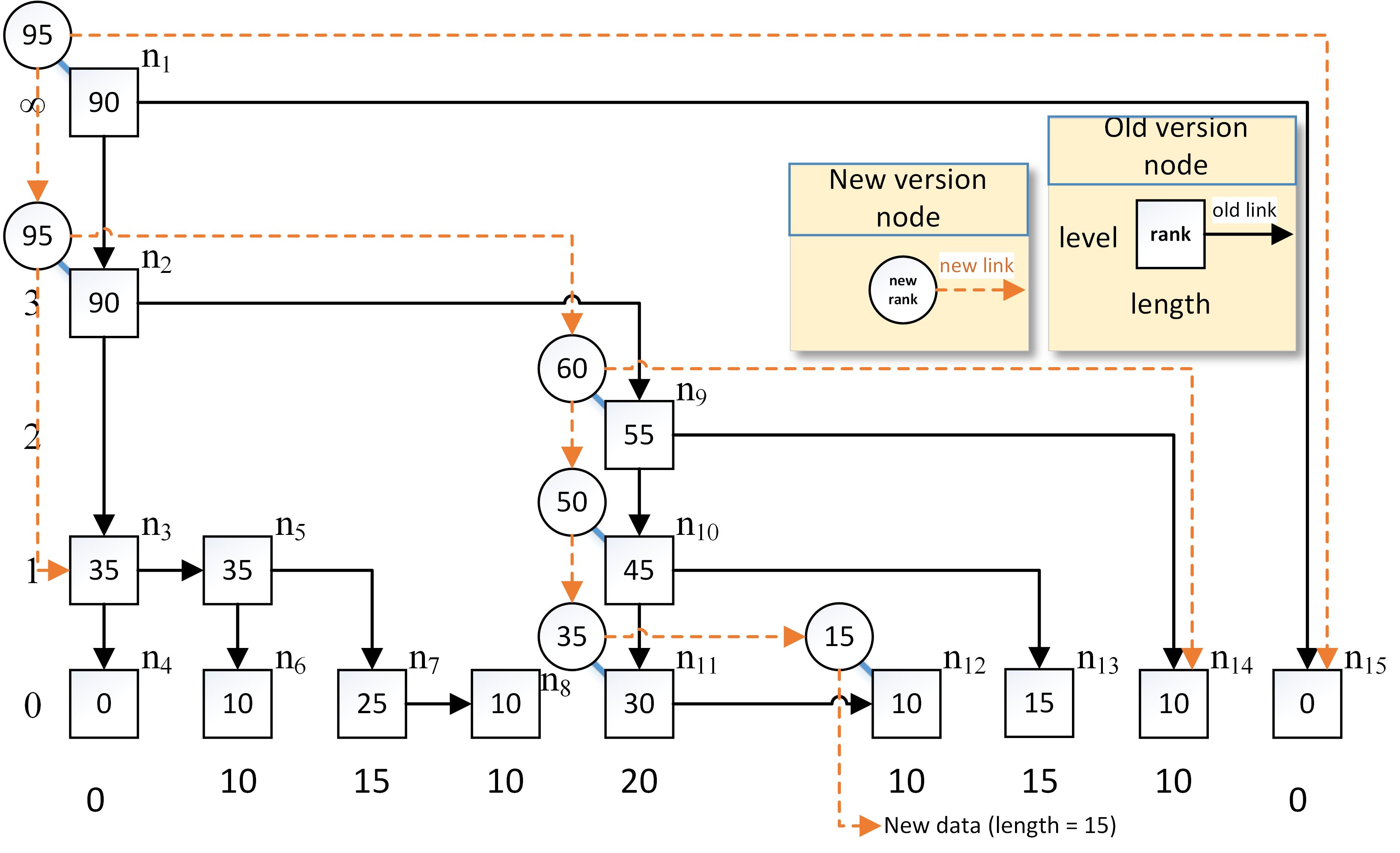}
	\caption{A \textit{modify} example on the \flexlist\ with versioning.}\label{fig:ToyModifyVersion}
	\end{center}
	\end{figure}
	
		We explain the modify algorithm (Algorithm \ref{alg:modify}) using Figure \ref{fig:ToyModifyVersion}. In the example the user modifies the block starting with the index 55 and puts a data block of length 15 there. We intend to keep the previous version as is and have a new root pointing to the new version of the data. Therefore we start by creating the new root using the old root's content for the next version (Line 1). Then we call the \textit{nextPos} with the new inputs. Version nodes, presented as circles in the figure, for n$_{1}$ and n$_{2}$ are created and attached as shown in the Figure. Line 7 and 8 creates version node for n$_{9}$ and links it. To finalize the modify algorithm, we loop until we reach the node to be modified by creating version nodes at each step (Line 9-12). We return two root nodes - one for the old version, one for the new version. Last step is to compute all the new rank and hash values from the stack for the new version. Since old version nodes remain untouched, we do not need any recalculation, and the same amount of calculations are needed as for the non-versioned auditable modify operation.

			\begin{algorithm}
		{\fontsize{\AlgFontSize}{\AlgSpace}\selectfont
		\SetKwComment{Comment}{}{}
		\KwIn{\Index}
		\KwOut{\CurrentNode, \Stack, \OldRoot}
		\BlankLine
		\Indp
		
			\underline{\textbf{\OldRoot\ = new Node(root, \Version)}}\;
			\underline{\textbf{\NewCurrentNode\ = \OldRoot}}\;
		    \Stack\ = new empty Stack \;
		    \CurrentNode\ $= root$ \;
		    call \NextPos\;
		    \CurrentNode\ = \CurrentNode.\After\;
		   \underline{\textbf{\NewCurrentNode.\After\ = new Node(\CurrentNode, \Version)}}\;
		   \underline{\textbf{\NewCurrentNode\ = \NewCurrentNode.\After\ then \NewCurrentNode\ is added to \Stack}}\;
		    \While{ \CurrentNode.\NodeLevel\ $ \ne 0 $ }{
		        \CurrentNode\ = \CurrentNode.\Below\ then \CurrentNode\ is added to \Stack\;
		        \underline{\textbf{\NewCurrentNode.\Below\ = new Node(\CurrentNode, \Version)}}\;
		        \underline{\textbf{\NewCurrentNode\ = \NewCurrentNode.\Below then \NewCurrentNode\ is added to \Stack}}\;
		    }
		    \underline{\textbf{save \OldRoot\ as the new root}}\;
		    \underline{\textbf{return $root$ as the previous version}}\; 
		\Indm
		\caption{Modify Algorithm}
		\label{alg:modify}
		}
		\end{algorithm}

	\subsubsection{Insert Operation}

		\begin{figure}
		\begin{center}
			\includegraphics[scale = 0.8]{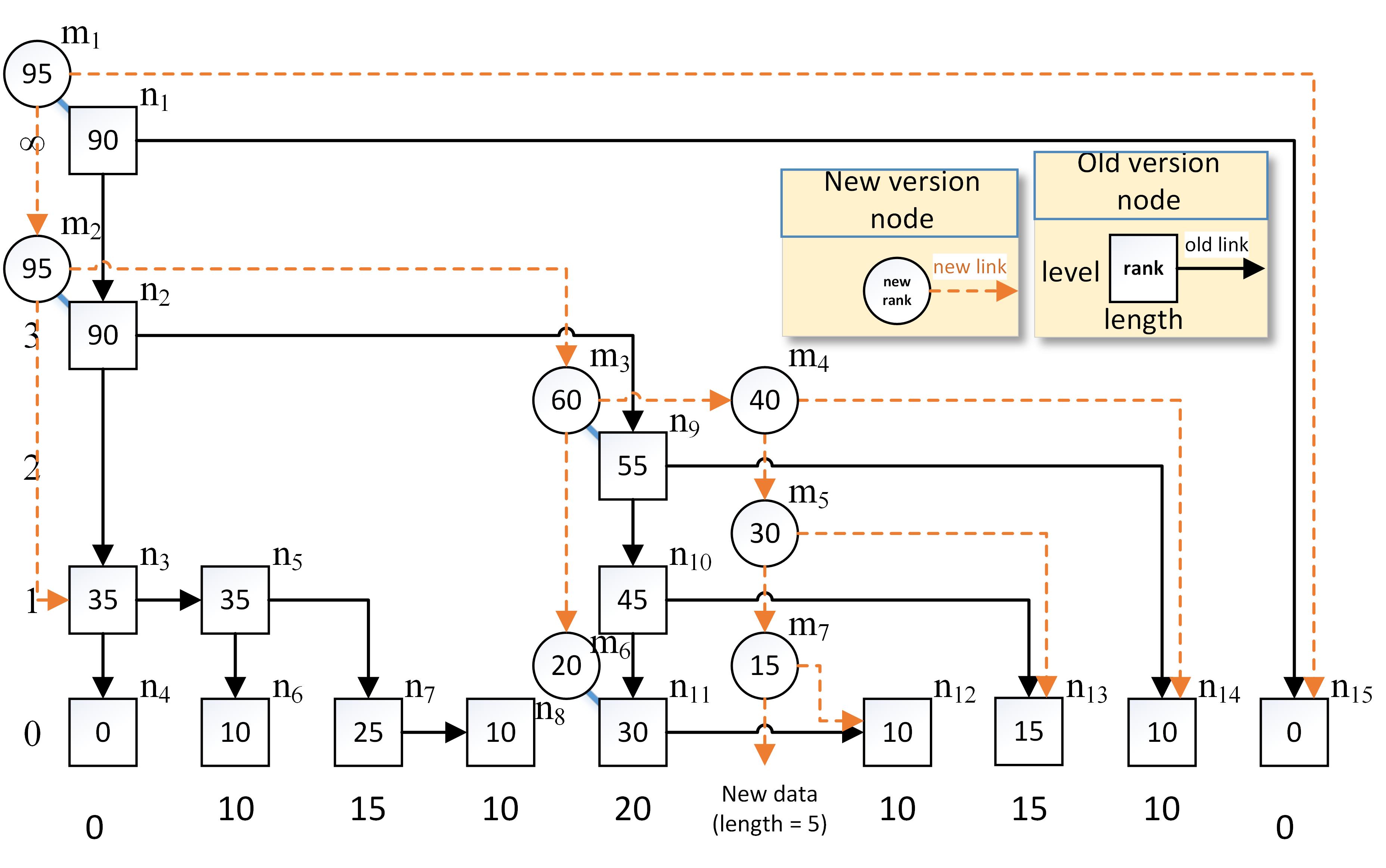}
		\caption{An \textit{insert} example on the \flexlist\ with versioning.}\label{fig:ToyInsertVersion}
		\end{center}
		\end{figure}
		
			\begin{algorithm}
		{\fontsize{\AlgFontSize}{\AlgSpace}\selectfont
		\SetKwComment{Comment}{}{}
		\KwIn{\Index, data}
		\KwOut{\NewNode,\Stack, \OldRoot}
		\BlankLine
		
		\Indp
			\underline{\textbf{\OldRoot\ = new Node(root, ++\Version)}}\;
			\underline{\textbf{\NewCurrentNode\ = \OldRoot}}\;
		    \Stack\ = new empty Stack \;
		    \PreviousNode\ $= root$; \CurrentNode\ $= root$; \Level\ $= tossCoins()$ \;
		    call \NextPos \Comment{ // \CurrentNode\ moves until it finds a missing node or \CurrentNode.\After\ is where \NewNode\ is to be inserted}
		    \If{$!CanGoBelow($\CurrentNode, \Index $)$ or \CurrentNode.\Level\ $\ne$ \Level}{
		        call $createMissingNode$; \Comment{// \textbf{instead of standart function call with \CurrentNode\ \underline{we call with \NewCurrentNode}}}
		        \underline{\textbf{\CurrentNode\ = \NewCurrentNode}}\;
		    }
		    \NewNode\ $=$  new node is created using \Level \;
		    \NewNode.\After\ = \CurrentNode.\After; \underline{\textbf{\NewCurrentNode.\After\ $=$ \NewNode}} and \NewNode\ is added to \Stack \;
		    \While{\CurrentNode.\Below\ $\ne null$}{
		        \If{\NewNode\ already has a non-empty after link}{
		            a new node is created to the \Below\ of \NewNode; \NewNode\ $=$ \NewNode.\Below\ and \NewNode\ is added to \Stack;
		        }
		        call \NextPos\; 
		        \NewNode.\After\ = \CurrentNode.\After; \NewNode.\Level\ = \CurrentNode.\Level \;
		        \underline{\textbf{$deteletUNode$(\PreviousNode, \NewCurrentNode)}}\;
		        \underline{\textbf{\CurrentNode\ = \NewCurrentNode\;}}
		    }
		
		\underline{\textbf{\NewNode.\LeafAfter\ = \CurrentNode.\LeafAfter}}\;
		\underline{\textbf{\NewCurrentNode.\LeafAfter\ = \NewNode}}\;
		    \NewNode $.data = data$;
		
			\underline{\textbf{save \OldRoot\ as the new root}}\;
			\underline{\textbf{return $root$ as the previous version}}\; 
		
		\Indm
		\caption{Insert Algorithm}
		 \label{alg:insert}
		}
		\end{algorithm}

	We explain the \textit{insert} algorithm (Algorithm \ref{alg:insert}) with Figure \ref{fig:ToyInsertVersion}. We repeat the same operation from Figure \ref{fig:ToyInsertFlexList} (insert data of length 5 to level 2) but we want to deduce two data structures with shared parts, as we do with the modify example. We start with creating the root m$_1$ for the new version. Then \textit{nextPos} moves the pointer until n$_9$ creating m$_2$ and m$_3$. Lines 6-8, a version node is created if it was necessary to connect newly created nodes (For example if the insertion was to level 3 instead of level 2, we would stop at n$_2$ and create a version node 
	after m$_2$ and above m$_3$). Then we create the new node of insertion, m$_4$, using the level to insert (Line 9) and set its after and m$_3$'s after. Lines 11-17 create m$_5$ and m$_7$ by taking their after links from n$_{10}$ and n$_{11}$. Line 16 deletes the version node for n$_{10}$ created by \textit{nextPos} since in the new version that node is unnecessary. Lines 18-19 connect the leaf nodes together. Then Line 20 sets the new data to the newly created node and the algorithm returns the old root for versioning. With the \textit{nextPos} pushing the newly created nodes to the stack, they re also pushed at Line 10 and 13. To finalize the new version we need to calculate rank and hash values of all the nodes from the stack.

	\subsubsection{Remove Operation}
	
			\begin{figure}[h!]
			\begin{center}
				\includegraphics[scale = 0.78]{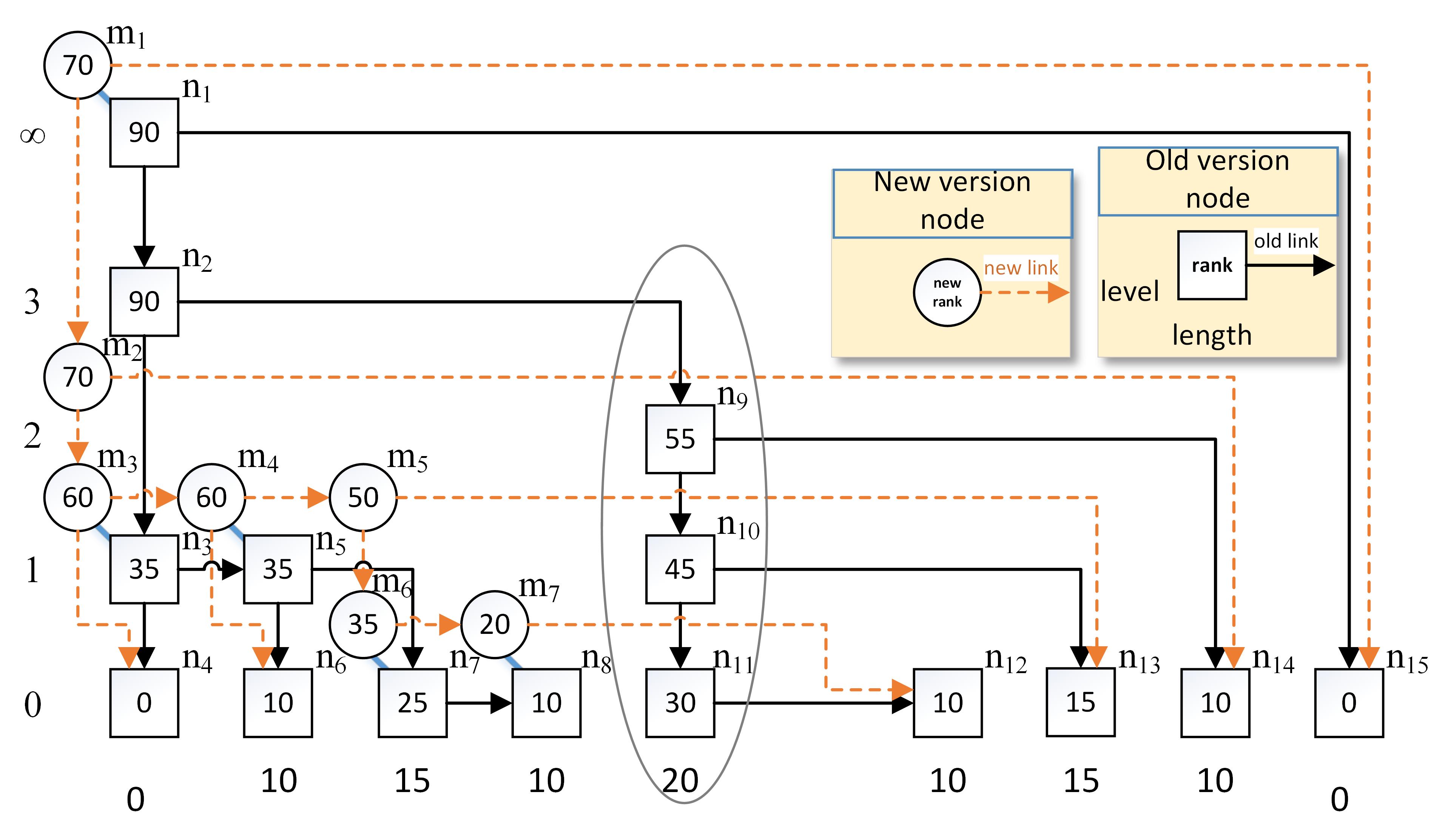}
			\caption{A \textit{remove} example on the \flexlist\ with versioning.}\label{fig:ToyRemoveVersion}
			\end{center}
			\end{figure}

	\begin{algorithm}
	{\fontsize{\AlgFontSize}{\AlgSpace}\selectfont
	\SetKwComment{Comment}{}{}
	\KwIn{\Index}
	\KwOut{\DeleteNode,\Stack, \OldRoot}
	\BlankLine
	
	\Indp
		\underline{\textbf{\OldRoot\ = new Node(root, ++\Version)}}\;
		\underline{\textbf{\NewCurrentNode\ = \OldRoot}}\;
	    \Stack\ = new empty Stack
	    \PreviousNode\ $= root$; \CurrentNode\ $= root$ \;
	    call \NextPos\; 
	    \DeleteNode\ = \CurrentNode.\After \;
	    \If{\CurrentNode.\Level\ = \DeleteNode.\Level}{
	    	\underline{\textbf{\NewCurrentNode.\After = \DeleteNode.\After}};
	    	\underline{\textbf{\DeleteNode\ = \DeleteNode.\Below}}; \Comment{ // unless at leaf level}
	    
	    }\Else{
	        $deleteUNode$(\PreviousNode, \NewCurrentNode)\;
	        \CurrentNode\ = \NewCurrentNode\;
	    }
	    \While{\CurrentNode.\Below\ $\ne$ null}{
	        call \NextPos\; 
	        
	        \underline{\textbf{call $createMissingNode$ with \NewCurrentNode instead of \CurrentNode}} \; 
	        \CurrentNode\ = \NewCurrentNode\;
	        \underline{\textbf{\NewCurrentNode.\After = \DeleteNode.\After}}; \Comment{// Deletion on the new version}
	        \DeleteNode\ = \DeleteNode.\Below\; 
	    }
	    	\underline{\textbf{\NewCurrentNode.\LeafAfter\ = \DeleteNode.\LeafAfter}}\;
	    	
	    	\underline{\textbf{save \OldRoot\ as the new root}}\;
	    	\underline{\textbf{return $root$ as the previous version}}\;

	\Indm
	\caption{Remove Algorithm}
	\label{alg:remove}
	}
	\end{algorithm}

	We run the remove algorithm (Algorithm \ref{alg:remove}) on our \flexlist\,, illustrated in Figure \ref{fig:ToyRemoveVersion} to remove the block starting with index 35. The new root, m$_1$ is created, and the \textit{nextPos} function brings the pointer to n$_2$ by creating a version node for n$_2$ as well. Since the level of the node to remove is not equal to the one our pointer is on (check at line 6) we delete the newly created version node and also pop it from the stack (called \textit{deleteUNode}). Then we create m$_2$ (the missing node - Line 13) since in the new version n$_{14}$ is connected to level 2 without the presence of n$_9$. After the linking of m$_2$ (Line 15) we let the \textit{nextPos} to move our pointer until n$_5$ by creating m$_3$ and m$_4$ on the way (line 12). At this point, we again need a node to connect the after link of n$_{10}$ therefore \textit{createMissingNode} (Line 13) creates one. Instead of creating the node on the \flexlist, we create and link it to the new version nodes. Therefore, m$_5$ is created and its after link is set to n$_{10}$'s after link (Line 15). The next call of the \textit{nextPos} function carries the pointer to n$_8$, creating m$_6$ and m$_7$ on the way. Then m$_7$ takes the after link of n$_{11}$ (the last of the nodes to delete). We set the leaf level links of the new version nodes and return the old root after setting the new one. \textit{createMissingNode} pushes the newly created node to the stack and the rest of the new nodes are pushed to the stack by the \textit{nextPos} function. To finalize the remove operation, all the hash and rank values of the nodes from the stack are calculated.

\subsection{Challenge Phase}

\subsubsection{Challenge}

Dynamic provable data possession systems employing an authenticated dictionary works in the following manner. The client, by sending a seed to the server sets the randomness for a shared pseudo randomness generator. Using that randomness, the server picks some blocks to generate a proof and collects all the necessary information to compute the hash value of the main root of the challenged data. Considering an authenticated dictionary, each node's hash value depends on the child nodes' hash values and their own attributes,  the necessary information are the attributes of the nodes in the path from the root to the challenged data blocks and their neighbor's hash values. The server also adds the data blocks themselves to the proof, so that the client may compute the root value using these unique values where one of them is the data block to be checked for integrity. Note that an authenticated dictionary uses a collision-resistant hash function, therefore these values are unique in the sense that there is no 
probabilistic polynomial time adversary that can find any other input sets to compute the same result. Formally: Let hash : $I_1$ x $I_2$ $\rightarrow$ O be a family of hash functions, a function is collision resistant if $\forall$ PPT adversaries A, $\exists$ a negligible function neg s.t. $P[( I_2, I_2')  \leftarrow A(I_1) : I_2 \neq I_2' \wedge ($hash$_{I_1} ( I_2) = $hash$_{I_1}(I_2'))] \leq neg(n)$.

We keep each of the data structures for each version so any version can be challenged in the
old way \cite{DPDP,blind}. To be precise, challenging the latest version as in \cite{DPDP,blind} schemes makes sense
but challenging all previous versions in that manner is too much of an overhead. So we suggest a version challenging mechanism to decrease this overhead, which challenges only the changed blocks from the versions. Since all versions are present in our scheme, additional computation in not needed, so, it is adequate to challenge only the changed nodes from each version, while other schemes (e.g., \cite{skipdeltaChen}) need to challenge the main version and the deltas until that version.

\subsubsection{Challenging Versions}

One solution would be to utilise the version numbers inside the nodes. We use an alternative that requires less authenticated information. We add a value to leaf level nodes of the layer 2 data structure (roots to layer 1), thus they keep the update information (index from which it starts, and the length) and add them to inputs of the hash function (authenticate) to compute the hash value for the node.


When we challenge the versions we choose particular nodes in a version that are modified in that specific version. To achieve that we use the authenticated information in the version's root to choose which nodes to challenge. Both parties use the seed that the client picks for a challenge and generate the indices to be challenged. Therefore the challenge by the client remains the same, no addition to original scheme \cite{blind} is needed. Since the client gets the start index and the length of the update by the proof from the layer 2, the server cannot choose a node other than whatever the client has picked by the random seed that she has chosen.

Thus, the proof consists of two parts. First is the proof from layer 2 which includes the roots of versions which are challenged. Second is the set of proofs for each version from layer 1. The integrity of the roots are guaranteed by layer 2 and the integrity of blocks are guaranteed by layer 1.  

Probability of catching a cheating prover when a fraction $f$ of a file is damaged or lost is $(1 - f)^r$ where $r$ is the number of blocks randomly picked and challenged \cite{PDP}. In case of loss or damage where $f =$ 10\% (of the versioning data), one needs 20 random nodes from the versions to be challenged to catch the problem with 88\%. Challenging 43 nodes is enough to catch the same problem with 99\%. Unlike in \cite{PDP,DPDP,scalablePDP} we don't consider 1\% of data loss since the versions, themselves, are small in size and deleting 1\% of it wouldn't benefit the server.

\subsection{Roll-back}
\label{sec:rollback}

Roll-back is achieved readily by choosing the root of the required version. Note that all indices of the version is at expected $O(logn)$ distance and recovering all of the version is $O(b)$ where $b$ is the number of blocks for that specific version. To be precise,  reaching the node at index 0 requires $O(h)$ hops, where $h$ is the height of the data structure, and following the leaf level nodes all the way right costs $O(b)$.
$O(h) = O(log(b))$ as discussed in Section \ref{text:insert}, therefore recovering all of the versions costs $O(log(b) + b)$.

Note that each algorithm that we provide in Section \ref{sec:ver} (line 19 of Algorithm \ref{alg:insert} and line 17 of Algorithm \ref{alg:remove}) deals with leaf level connections which are not presented in the Figures for simplicity.

\subsection{Git Integration}

Git is a popular versioning system because it achieves very good performance given the underlying data model for storing versioned data. Even though our approach can be plugged into many other versioning systems, we choose Git as the candidate to demonstrate the proposed approach's applicability, both because Git is a widely used system with its superior performance than the delta based solutions as well as because our approach naturally complements Git's underlying data model, and enables the use of much less storage space for each version (than Git), thus scaling the size of data objects that can be supported, while retaining the same efficiency of retrieving arbitrary versions of data.

\begin{figure}[htb]
\begin{center}
	\includegraphics[width = 1\columnwidth]{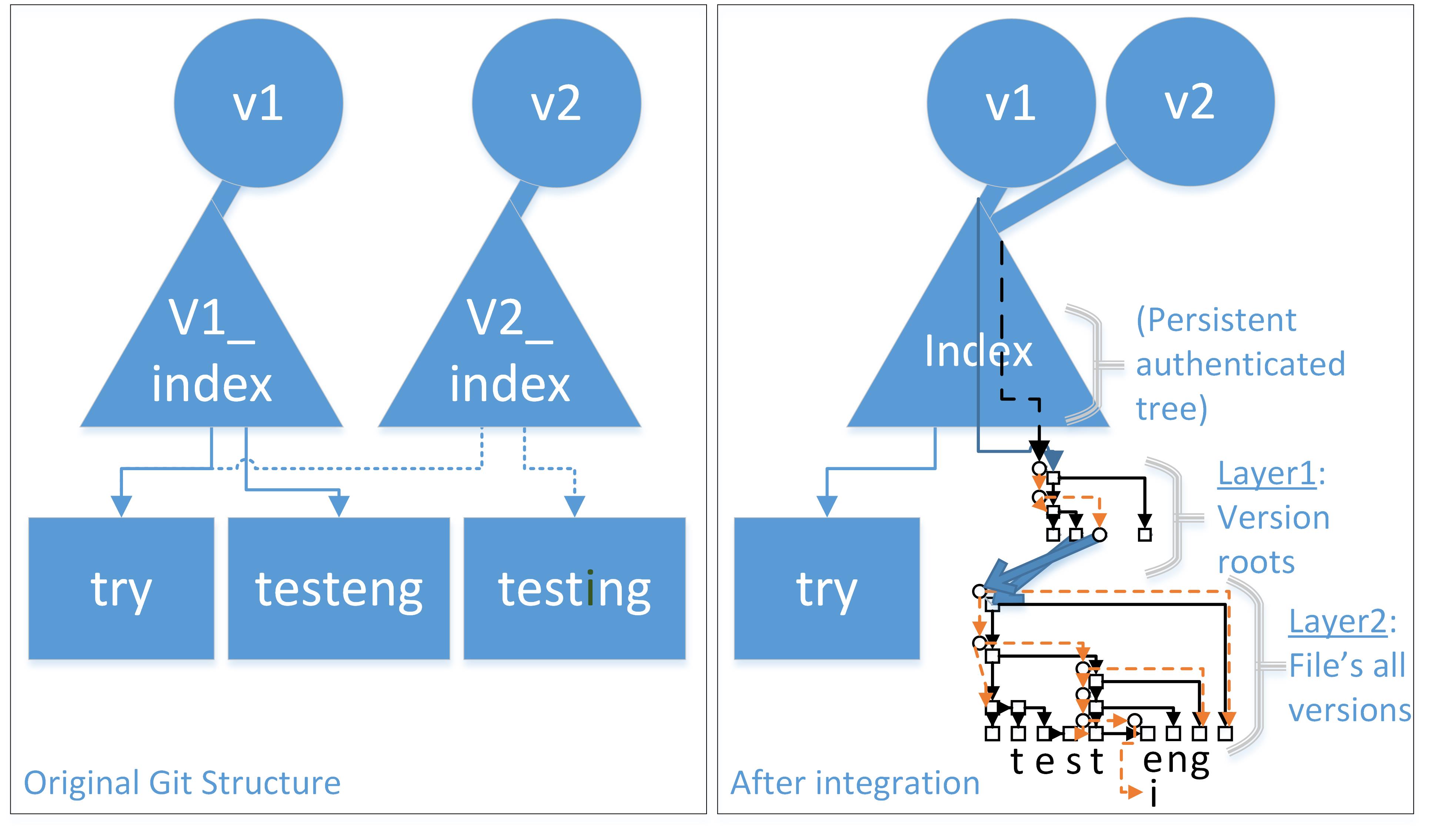}
\caption{How Git would look like after integration.}
\label{fig:gitintegration}
\end{center}
\end{figure}

\begin{figure}
\begin{center}
 \includegraphics[width=0.9\textwidth]{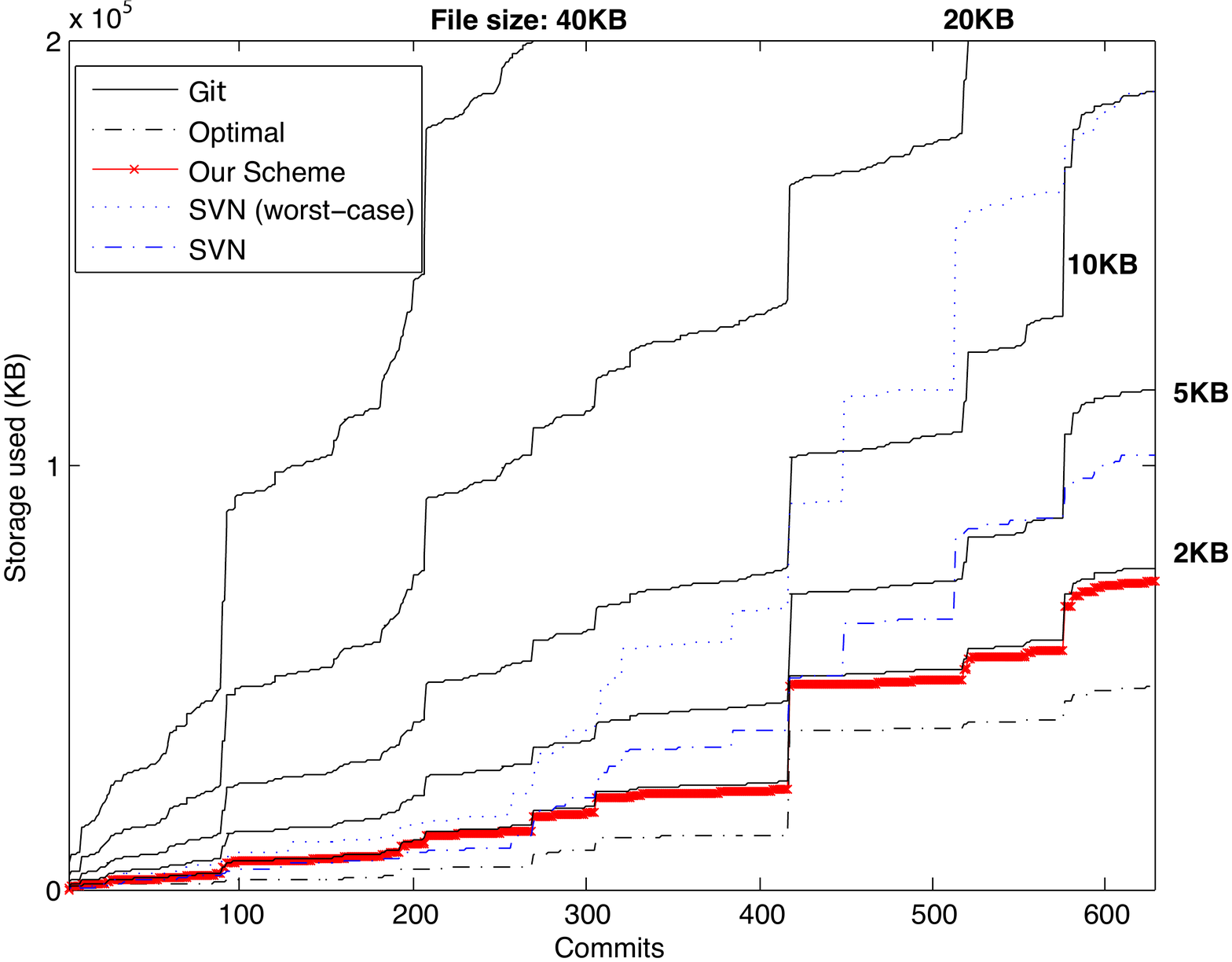}
                \caption{Space requirement after 627 commits.}
                \label{fig:optimal_persistance_git}
 \end{center}
\end{figure}

In Figure \ref{fig:gitintegration} we show how our approach can be integrated with and supplements Git \cite{gitt}. On the left we show how Git works. In this example, a typo is fixed by changing a character in a file. Git creates a new blob for that file and adds it to the index where it keeps an index tree for each version. On the right, we show how our approach can be integrated to supplement Git. We suggest usage of an persistent (to save space by not repeating shared node's hash values) authenticated tree structure for indexing (e.g., a Merkle hash tree \cite{merkleHASH} would suffice, since we don't need re-balancing at this layer). When we use our proposed solution, shown on the right (with logical block size of 1 in this example), instead of creating a new blob, only the changed part is subject to change. Following the index starting from the root shown by `v2' the data we will reach is the version 2. No further de/compression is required.

This example demonstrates that our approach deals with the data at a different granularity. While Git provides blob level deduplication \drafta{(which is a slightly better way than file-level deduplication, where the deduplication excludes the file header and focus on the content itself where two files with, for instance, different authors, name or date can be deduplicated), our deduplication approach becomes active when the file is updated below the blob level, where we handle deduplication within a blob at the block level.}
For a file of size 400KB, 5 alterations of average size of 20KB will end up with a space usage of $\sim$2000KB with Git (for storing the five versions), on the other hand, integration of our approach will cost 100KB in addition to the 400KB, using a total space of 500KB. We note that Git applies compression (zlib \cite{gitt}) to reduce the storage overhead, and the numbers we report in this example are for just a qualitative comparison. Furthermore, applying compression induces further computational costs. Overall, the advantages of our approach are directly proportional to the size of the files being changed. 

In Figure \ref{fig:optimal_persistance_git}, we show the space usage at the server for a mutable data. In this real life example deducted from our own SVN server, we track a file directory that starts with several files of total size 117KB. At the end of 627 commits the size gets increased to 50MB. By taking these commits as basis, we show the space requirement for the optimal case, where only the necessary bytes are kept. This is the optimal, where, even the delta based solutions (such as SVN \drafta{without the skip delta implementation}) do not achieve but approximate. \drafta{We show the SVN space usage while using the skip deltas. In the worst case, all updates are additive and in the average case the updates are as they really are}. We present the Git results for different average file sizes (2KB, 5KB, 10KB, 20KB, 40KB). We observe that even when the average file size is 2KB, Git uses more space than our variable block size setting, since Git copies the same file and adds the additional content on it, where we may add the additional content as a new block. The bigger the average file size, the more space is used in Git. With our solution integrated to Git, regardless of the average file size, the overhead depends on the updates only.

\section{Evaluation}\label{sec:eval}

We have employed methods from cashlib library to implement the proposed mechanisms \cite{cashlib}. The experiments are run on a 64-bit computer with a 3.2GHz Intel Xeon E5-1650 Processor with one active core, \drafta{Dell Perc H310 (for high density, entry level servers) disk drive that provides 6Gb/s per port}, 15.6GB main memory and 12MB L2 cache, running Ubuntu 12.10. As security parameters, for a 80-bits expected security we use 1024-bit RSA modulus, 80-bit random numbers, and SHA-1 hash function. Reported results are the averages of 10 iteration of experiments. The tests include I/O access time. The \flexlist\ was small enough to be stored in the main memory. Unless stated otherwise, \textbf{we use data of size 1GB in our experiments}.

There is a dearth of suitable data sets (see a discussion in the context of deduplication \cite{commitpettern}) to drive the experiments. We thus created synthetic commit patterns, generated randomly - parametrized by ranges of sizes of updates, which allows the study of a wide range of update behaviors. Specifically, we use a uniform random distribution and the updates are insertion, removal or combination of both. We choose ranges as follows; 1KB to 10KB, 10KB to 20KB, 20KB to 50KB and 50KB to 100KB. We use 10KB to 50KB range to show time overheads since they are relatively big, and allows to better observe latency, and we want to demonstrate that the overheads are acceptable even for such moderately large updates. We used 1KB to 20KB range to show space overheads since an average update size falls into this range for many scenarios \cite{DPDP,commitpettern}.

\begin{figure}[!ht]
        \centering
        \begin{subfigure}[b]{0.47\textwidth}
		                \includegraphics[width=1.1\textwidth]{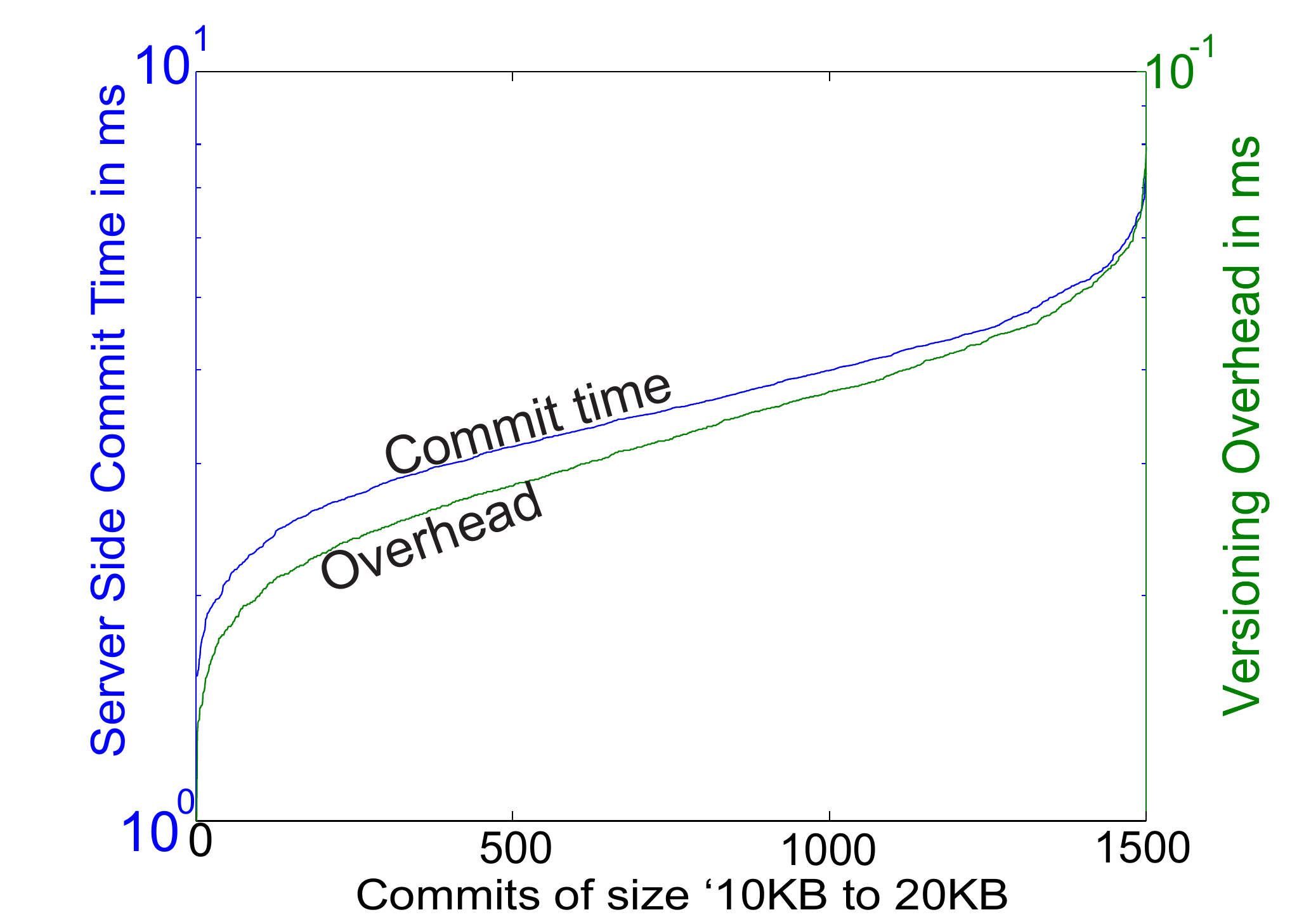}
		                    \caption{Time spent for each of 1500 commits of size 10KB to 20KB  on 1GB file with block size 2KB, sorted for better visualization.}
		                    \label{fig:VersioningOverheads1020}
        \end{subfigure}%
        ~ 
           \quad
        \begin{subfigure}[b]{0.47\textwidth}
		                \includegraphics[width=1.1\textwidth]{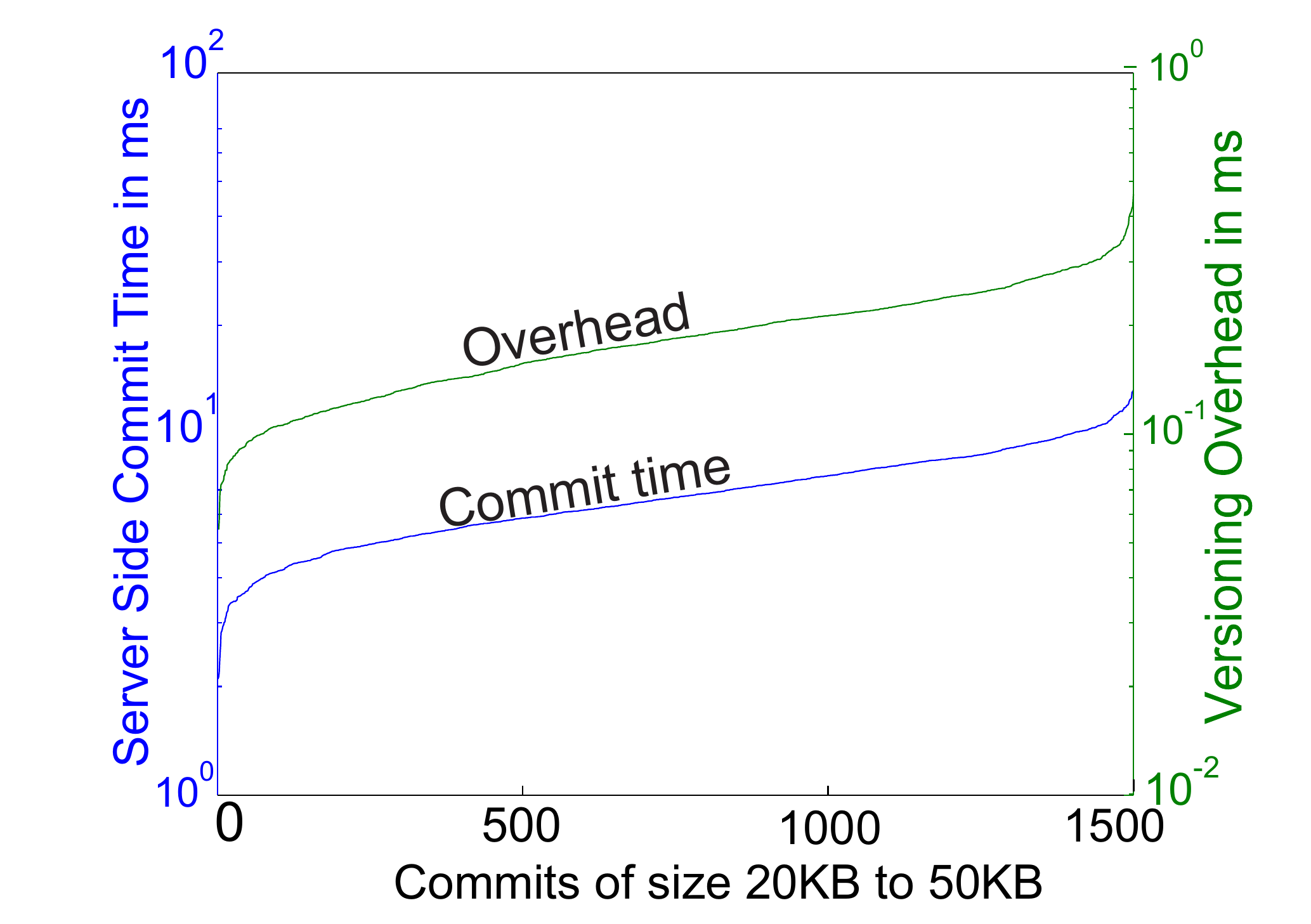}
		                \caption{Time spent for each of 1500 commits of size 20KB to 50KB on 1GB file with block size 2KB, sorted for better visualization.}
		                 \label{fig:VersioningOverheads2050}
        \end{subfigure}

        \caption{Time spent for commits.}\label{fig:OverHeadsByKBcommitsize}
\end{figure}

The overhead in terms of time spent for versioning with respect to \cite{blind,DPDP} is presented in Figures \ref{fig:VersioningOverheads1020} and \ref{fig:VersioningOverheads2050}. The versioning overhead is around 5\% and 4\% for updates ranging from 10KB to 20KB and 20KB to 50KB respectively. We see that, among 1500 commits, some with big size and including operations to higher levels of the data structures ends up with significantly higher overheads (shown on the rightmost of the graphs). For instance, an update of size 50KB (25 blocks involved, where 17 of them are inserts, among which the maximum level is 5 and average level is close to 1) takes less than 18ms for the commit operation and 0.4ms  as the versioning overhead. Regarding these events with significantly higher overheads, we note that even though an update takes long time, the versioning overhead is barely visible. Not only it is too small in the update phase, but also with the proofs and server-client communication added (which ends up taking seconds) the versioning overhead is negligible.

\begin{figure}[!ht]
        \centering
        \begin{subfigure}[b]{0.47\textwidth}
		                    \includegraphics[width=1.1\textwidth]{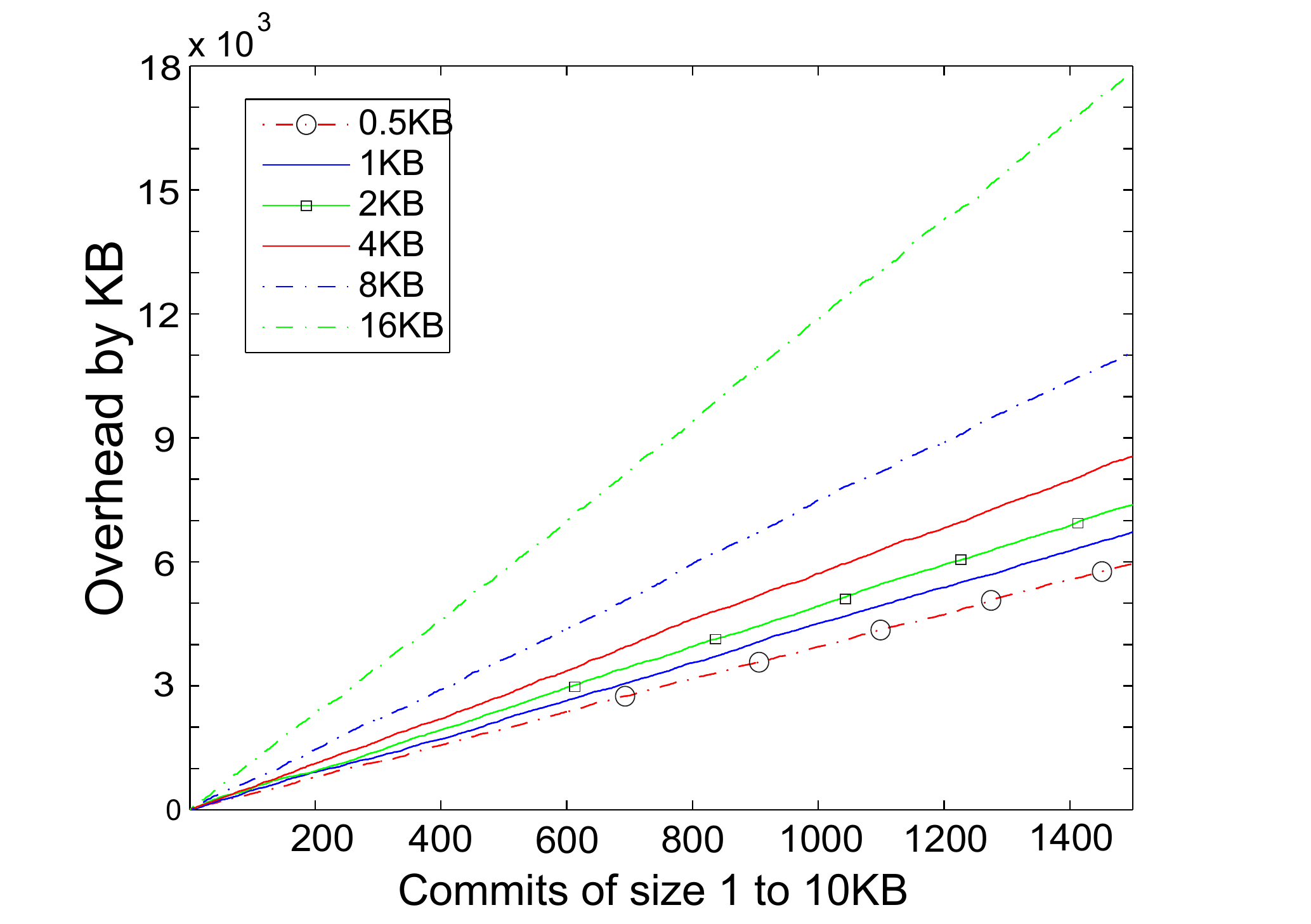}
		                     \caption{Space overhead of versioning 1KB to 10KB  on 1GB file with different block sizes.}
		                     \label{fig:OverHeadsByKBcommitsize0110}
        \end{subfigure}%
        ~ 
           \quad
        \begin{subfigure}[b]{0.47\textwidth}
		                    \includegraphics[width=1.1\textwidth]{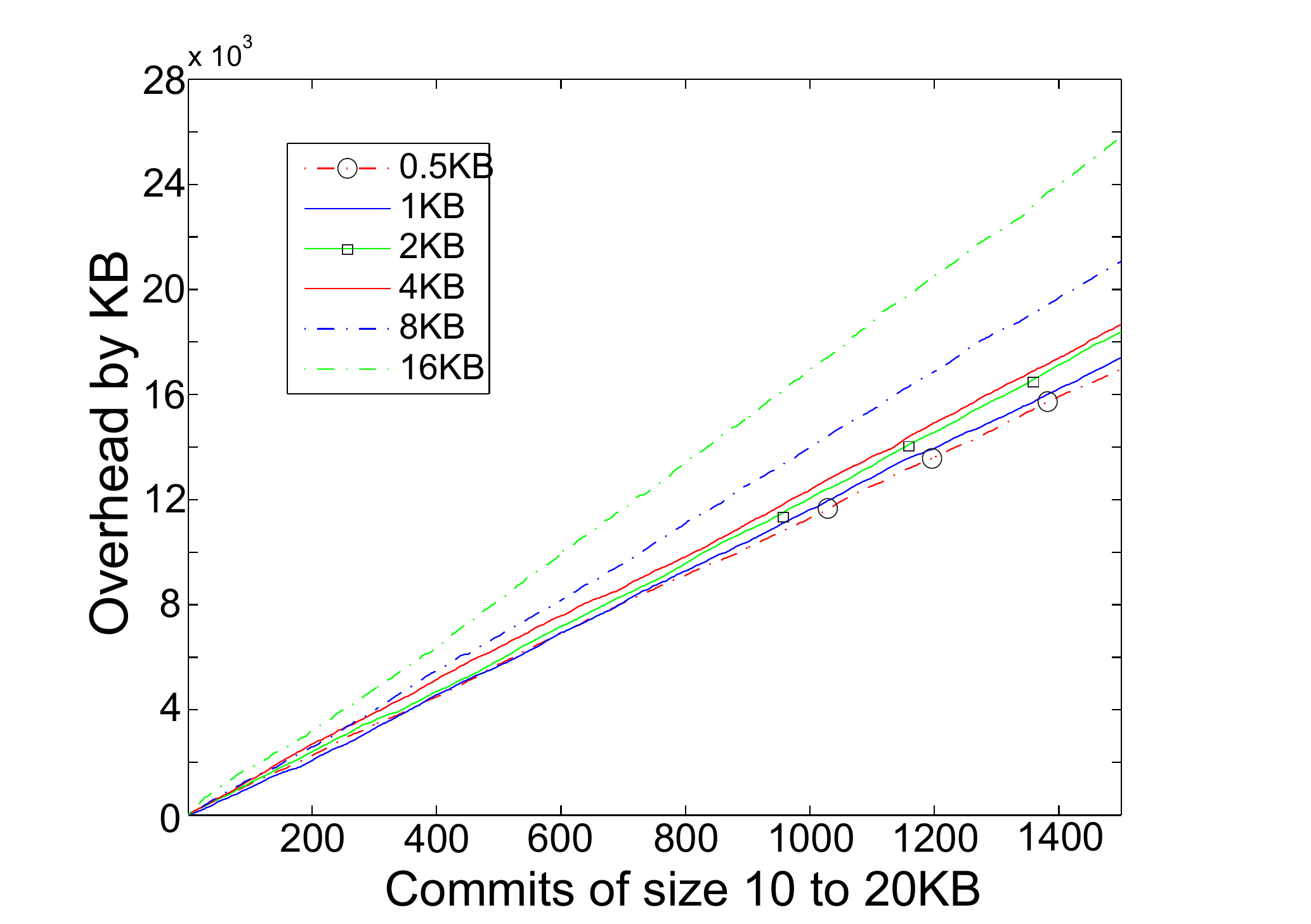}
		                   \caption{Space overhead of versioning 10KB to 20KB  on 1GB file with different block sizes.}
		                     \label{fig:OverHeadsByKBcommitsize1020}
        \end{subfigure}

        \caption{Space overhead of versioning.}\label{fig:SpaceOverHeadsByKBcommitsize}
\end{figure}

Figure \ref{fig:OverHeadsByKBcommitsize0110} and \ref{fig:OverHeadsByKBcommitsize1020} show the different (due to the block size choice) overheads to keep previous versions of the data. In \cite{blind}, it is shown that 2 and 4KB block sizes give best performance results for auditing. With versioning (this work), 2-4KB block sizes have reasonable space usage on the server side (not much is gained with smaller blocks of say 0.5KB), thus, there are no conflicts in choosing the design parameters.

\begin{figure}[!ht] 
\begin{center}
                \includegraphics[width=0.8\textwidth]{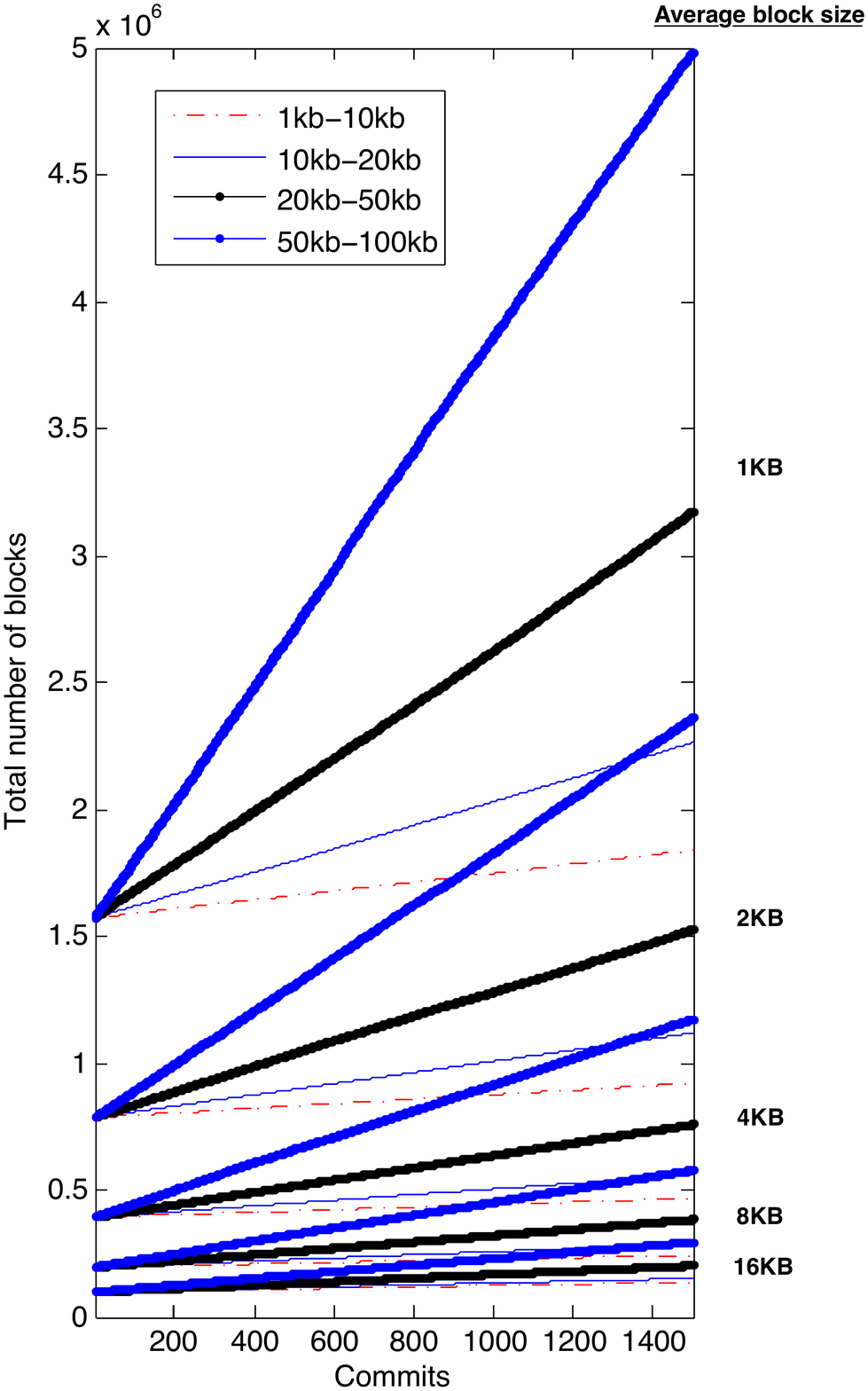}
                \caption{All blocks of the FlexList for 1500 commits of different sizes on 1GB file with different block sizes.}
                \label{fig:ExtraBlocksOnTheFlexListMarked}
 \end{center}
 \vspace{-10pt}
\end{figure}

Figure \ref{fig:ExtraBlocksOnTheFlexListMarked} shows the increase of the total number of nodes (including all versions) in the \flexlist\ under different settings. For the most common update sizes of 1 to 10KB range, the increase in node number is barely discernible. If big commits of sizes 20 to 50KB arrive for consecutive 1500 commits (which is unlikely) while using a 2KB block sized \flexlist , number of the nodes doubles. Each node constitutes of 5 integers and one 160 bit hash value (considering SHA-1 as the hash function) and leaf level nodes includes extra 1024bit tags for network efficiency ($\sim$60MB for  1GB of data). Therefore doubling the \flexlist\ means using an extra space of $\sim$76MB for the data structure if an integer is considered to be of 2 bytes. The smaller the block size per node, the less overhead, while when it gets bigger, FlexList loses some efficiency. We see that the persistent \flexlist\ and our 2KB choice stays in an acceptable range in terms of extra nodes used and space consumed. If under any setting the $\sim$76MB space is not affordable, then larger block sizes may be used at the cost of processing time.
For instance, when dealing with a big volume of data but with only insertion based modifications, bigger block sizes could be used.

\begin{figure}[!ht]
\begin{center}
                \includegraphics[width=\textwidth]{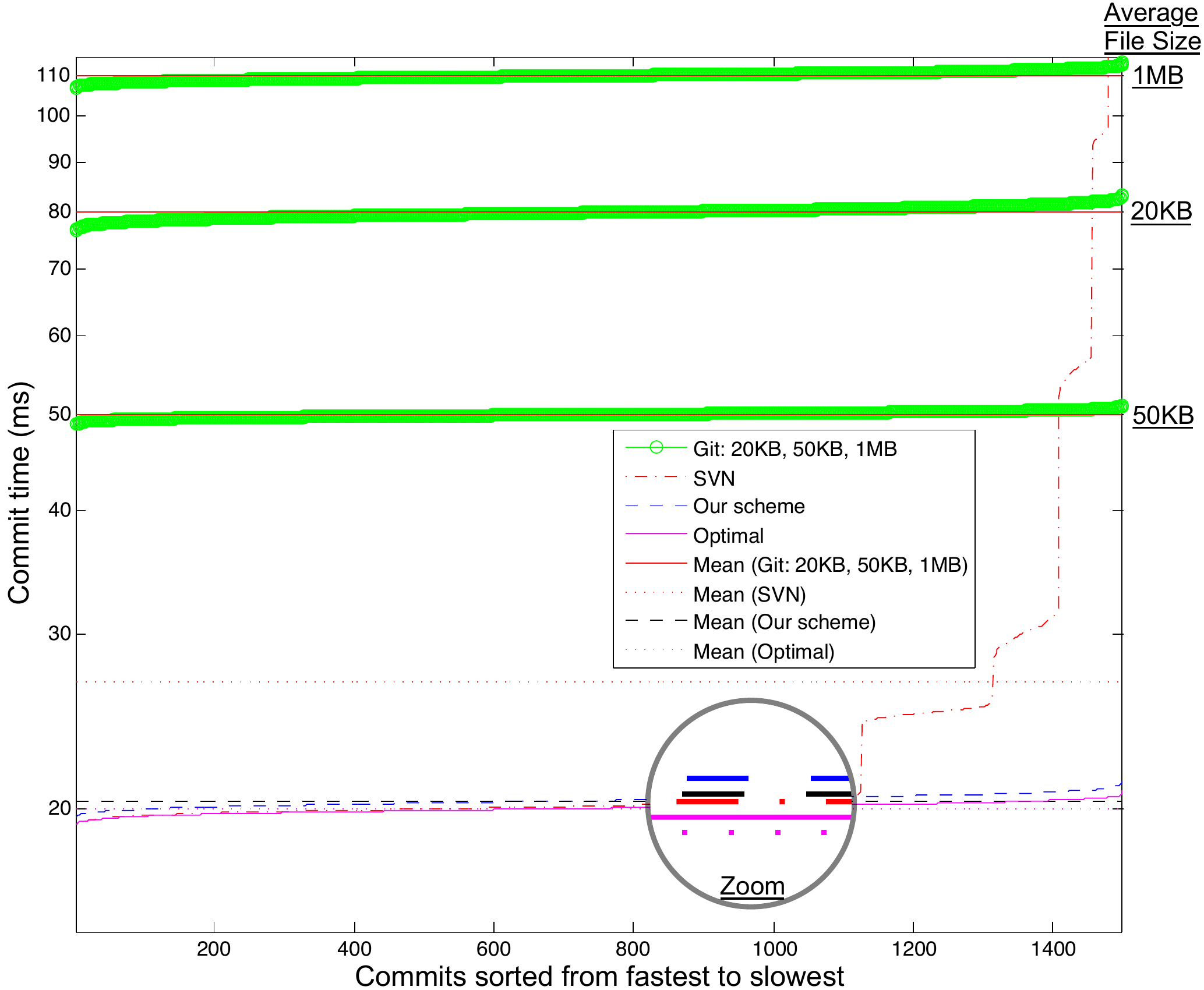}
                \caption{Server Side Commit times.}
                \label{fig:CommitTimesSVNGitFlex_after2ndReview}
 \end{center}
\end{figure}

\drafta{Integrating our scheme with Git is logically straightforward, however, ultimately, substantial software integration work. In order to determine the advantages of our approach, we instead emulated the Git part and tested our implementation as another module working on top of Git. Figure \ref{fig:CommitTimesSVNGitFlex_after2ndReview} shows the time spent at the server side for each commit. We assume a 1GB data stored. We applied 1500 randomly generated commits. Each commit size is between 20KB and 50KB. We sorted the commit times from the fastest to slowest for better visualization. For the optimal case we created a delta that includes only the differentiated part of the file and measure the time. The optimal is not a result of an actual scheme but measured only as a base reference. For Git, we have chosen 3 average file sizes of 20KB, 50KB and 1MB. In the legend of Figure \ref{fig:CommitTimesSVNGitFlex_after2ndReview}, we have grouped these together into one. When the file sizes diminish, the file hierarchy grows, resulting is a bigger tree structure (file hierarchy). Since the tree structure is copied with each update, the effect is more pronounced with the smaller file sizes. On the other hand, when the average file size is 1MB, since the file itself is bigger than the space used by the whole file hierarchy, it dominates the time spent for the commit. For SVN, we observe an exponential increase in time spent due to the size growth of the skip deltas with the increase in the commit count. Even with a very fast storage device (Dell Perc H310 for high density, entry level servers disk drive that provides 6Gb/s per port), we observe that the main factor in the spent time is the disk write times of the files, deltas or the data structures. Our scheme (applied along with Git) minimizes the data writes and does the updates in variable sized blocks and does not copy the whole file hierarchy, thus the difference from one version to another is close to optimal data writes. Regardless of the average file size, it makes all commit times in Git with our scheme quite close to the optimal solution with an average 2KB block sizes.}

\drafta{Last but not the least, as we have discussed in Section \ref{sec:rollback}, one of the best features in our scheme is that it allows an instantaneous roll-back to any version (like the original Git) without any processing required. The only thing the server needs to do is to choose the root of the particular version (as the root of the file hierarchy) and all the leaf level nodes constitute the data of that version. Garbage collection of redundant nodes/links, if desired by a user, can be carried out in a disentangled manner as a background process, and does not affect the run-time performance.}

\section{Conclusion}\label{sec:conc}
In this paper we have proposed a practical solution for auditing versioned data, supporting arbitrary edits across versions, while not compromising performance of the various operations (access/commit/audit). We have discussed how the proposed mechanism supplements version control systems like Git providing significant storage savings in addition to auditability.

\section*{Acknowledgements}\label{sec:ack}

Ertem Esiner's research was financially supported by the Singapore International Graduate Award (SINGA), grant no. TU-2013-R13145800000000. We would like to thank Prof. Dr. Alptekin K\"up\c{c}\"u and Mohammad Etemad for their constructive inputs in course of this work.

\section*{References}

\bibliographystyle{elsarticle-num}

\bibliography{ertem}



\end{document}